\documentclass[default,iicol]{sn-jnl}
\usepackage{filecontents}
\usepackage{stfloats}
\usepackage{amsmath,amssymb,amsfonts}
\usepackage{graphicx}
\usepackage{textcomp}
\usepackage{moreverb,url}
\usepackage[utf8]{inputenc}
\usepackage[english]{babel}
\usepackage{xcolor}
\usepackage{soul}
\usepackage{lineno,hyperref}
\usepackage[caption=false,font=footnotesize]{subfig}
\modulolinenumbers[5]
\usepackage{siunitx}
\usepackage{float}
\usepackage{caption}
\usepackage{lscape}
\usepackage{rotating}
\usepackage{mathtools}
\usepackage{arydshln}
\usepackage{multirow}
\usepackage{wasysym}
\usepackage{yfonts}
\usepackage{ae,aecompl}
\usepackage{blindtext}
\usepackage{soul}
\usepackage{ragged2e}
\usepackage{etoolbox}
\usepackage{lipsum}
 \usepackage{sidecap}
 \usepackage{mathptmx}
 \usepackage{mathpazo}
 \usepackage{optidef}
 \usepackage{setspace}
 \usepackage{color}
 \usepackage{enumerate}
 \usepackage{multimedia}
 \usepackage{ulem,lipsum}
 \modulolinenumbers[5]
 \usepackage{enumitem}
\begin{document}

\title[Article Title]{Linear matrix inequality based Type-III compensator synthesis for DC-DC converters}

\author*[1]{\fnm{Rıdvan} \sur{Keskin}}\email{ridvan.keskin@beun.edu.tr}

\author[2]{\fnm{Ibrahim} \sur{Aliskan}}\email{ialiskan@gmail.com}

\affil*[1]{\orgdiv{Department of Electrical and Electronics Engineering}, \orgname{Zonguldak Bülent Ecevit University}, \orgaddress{ \city{Zonguldak}, \postcode{67100}, \country{Türkiye}}}

\affil[2]{\orgdiv{Department of Control and Automation Engineering}, \orgname{Yildiz Technical University}, \orgaddress{ \city{İstanbul}, \postcode{34220}, \country{Türkiye}}}

\abstract{Boost, buck-boost, and fly-back DC-DC converters which are utilized in power lines of any electric vehicles, solar energy, and power factor correction applications require control systems to regulate the output voltage under mismatched disturbances i.e. load current and input voltage. In continuous current mode operation, the converters, however, are bandwidth-limited control systems due to their non-minimum phase nature. Disturbance rejection performance of such bandwidth-limited control system is an open problem especially where input voltage and load current disturbances cannot be measured. A third-order integral-lead (Type-III) compensator with a disturbance observer (DOB) can suppress the disturbances and unmodeled dynamics of the converters. However, synthesizing such a fixed-order control system under performance constraints is generally challenging. This paper proposes a simultaneous design of a Type-III compensator and a fixed order DOB based on $H_\infty$ control approach using convex optimization. The optimization problem is formulated in a convex-concave procedure by including the estimated disturbance and sensor noise functions. We proposed a  two-stage iterative algorithm to solve the problem in a convex optimization framework. Convex programming can therefore be used to synthesize an optimal fixed-order control system by removing the non-convex constraints on the parameter space. The approach leads to an easily resolvable control algorithm with linear matrix inequality constraints over parameterized controller parameters due to the convexity of the problem. The proposed control system is implemented on a 200 W DC-DC multi-phase interleaved boost converter prototype using a TMS320F28335 digital signal processor. The performance of the approach is compared with the well-known K-factor design approach for the Type-III compensators.}

\maketitle

\section{Introduction}\label{sec1}
Boost, buck-boost, and fly-back converters are utilized in power transmission lines of any electric vehicle, solar energy systems, and high voltage DC-DC applications \cite{azarastemal2021cascade}. The converters are improved to operate in high-power applications through any kind of parallel, cascade, and interleaved techniques. The converters regulate the output voltage of the converter-connected systems in the presence of load current and line variations. Linearization of the nonlinear converter models around an operating point produces an unstable zero which causes additional phase lag, extreme overshoots, and unpleasant responses in a set-point step change in the voltage mode control loop.  Therefore, synthesizing a voltage mode compensator for the converters is challenging \cite{shruti2021analytical}. In large part of control loops in industrial applications, popular fixed-order controllers like PI, PID, lead, lag, lead-lag, and third-order integral-lead (Type-III) compensators are preferred to overcome such issues. The Type-III compensator, a special kind of power electronics compensator, provides solid disturbance rejection capability due to inherent phase boost to maintain a reasonable phase margin \cite{sarrafan2020novel}. Unlike advanced robust PI, PD/PID, and lead-lag controller synthesize methods commonly exist in the literature \cite{hayes2016design}, there exist control design methods that consider only nominal operating conditions for Type-III compensators. 

The K-factor method which depends on the designer's prior control knowledge is widely used for designing the compensator \cite{chen2020interleaved, ghosh2016design, rana2017development, anzehaee2018augmenting}. However, the design process generally does not consider unavoidable noise, disturbances, and uncertainties. The Pole-placement method is proposed for designing the compensator according to ideal closed-loop models \cite{keskin2019design}. However, the method is derived from the designer's prior control knowledge of engineers \cite{cr2011type}. Machine-learning or swarm algorithms, for example, particle swarm optimization \cite{banerjee2016improved}, genetic algorithm \cite{tran2020dual} and dragonfly algorithm \cite{rana2021performance} are used to optimize the parameters of the compensator considering only the step response of the control system. The algorithms generally converge to local solutions since they are non-convex optimization algorithms. The stopping criteria used for the global solution in such optimization algorithms are usually random and vary according to the lower and upper bounds of the controller parameters \cite{hindi2004tutorial}. A robust non-convex $H_\infty$ method which considers the sensor noise and parameter uncertainties in the design process is proposed to design a Type-III compensator \cite{keskin2021robust}. The parameters of the compensator obtained from resolving the algorithm are different since the non-convex algorithms generally converge to different local solutions in each run of execution. These methods may be insufficient for industrial applications since the methods do not involve sensor noise, parameter uncertainty, and mismatched disturbances in the design process \cite{sedhom2020robust}. Robust $H_\infty$ structured control techniques guarantee robust stability and performance challenges, particularly the approaches based on linear matrix inequalities (LMIs) \cite{bagheri2022robust}. A fixed-order controller synthesis is restructured as a constrained convex optimization problem where the local solution is also the global solution \cite{ahmad2017robust}. Youla parameterization \cite{rodriguez2020modeling}, convex inner approximation \cite{erol2019fixed}, augmented Lagrangian \cite{ankelhed2012partially} and convex-concave procedure (CCP) \cite{kammer2017decentralized} approaches are utilized to convert the problem into a constrained convex form. A convex optimization method which includes converting bilinear matrix inequality constraints to LMI constraints is proposed \cite{dinh2011combining}. It is presented that the convex-concave algorithm converges to a locally optimal solution. A controller synthesis method is proposed, where the method covers only fixed-order controllers with parameters to be optimized only in the numerator matrix, such as PI and PID \cite{saeki2010low}. Another PID controller synthesis method is presented to assess mutual interference via frequency response data \cite{shinoda2017multivariable}. A PID controller design method whose time constant is fixed at a pre-defined value is proposed using convex optimization \cite{boyd2016mimo}. The parameter uncertainty of the system is represented in the inner loop as mini circles of the loop transfer function. The time constant of the PID controller can be optimized iteratively \cite{segovia2013noise}. Quantitative feedback theory-based automatic loop-shaping method is proposed for PID controllers in multi-model uncertain systems \cite{mercader2016robust}. \par 
The aforementioned methods can only be used for controller types that have parameters to be optimized in the numerator polynomial. The Type-III compensator design problem, however, is inherently non-convex since it has parameters to be optimized in the denominator polynomial. A PI controller with input filters is synthesized using a frequency-domain approach for multiple voltage source inverters \cite{kammer2018convex}. The approach allows designers to use $H_2$, $H_\infty$, and $H_2-H_\infty$ loop-shaping performance objectives. A controller synthesis approach is proposed for model-based or data-driven systems \cite{karimi2017data}. \par
A disturbance observer (DOB) theory deals with online unmodelled system dynamics estimation and compensation, and it consists of an inverse of the nominal system model and a typical low-pass filter ($Q$) for the system which has input disturbances \cite{ohnishi1996motion}. The filter characterizes the transient response and uncertainty rejection performances of the DOB. The various DOB structures and design methods are summarized \cite{sariyildiz2019disturbance}. An LMI-based control method for designing $Q$-filter is proposed assuming the control model is a minimum phase system \cite{wang2004design}. A systematic DOB design methodology for non-minimum phase systems is proposed in \cite{wang2020robust}. The methodology considers non-convex constraints for the internal stability of the feedback controller and DOB. Another non-convex design method of DOB is proposed, where the fixed-order $Q$-filter consists of a single tuning parameter \cite{tena2022performance}. A sequential design method of a DOB and full-order feedback controller is proposed using the non-convex $H_\infty$ mixed-sensitivity approach \cite{huang2022high}. A simultaneous design of inverse of the plant and DOB is proposed for data-driven systems in an $H_\infty$ convex optimization framework, where the feedback controller is pre-designed \cite{wang2022frequency}. The mentioned studies are implemented for the systems which have input disturbances. A genetic algorithm-based DOB design method is proposed for the systems that have mismatched disturbances, where the inverse of the plant model is estimated through the concept of equivalent transfer function \cite{gonccalves2022disturbance}. We propose an approach in which the fixed-order feedback controller and the $Q$-filter are synthesized simultaneously in a convex optimization framework for control systems that have mismatched disturbances.

In this paper, a simultaneous controller design method for the Type-III compensator and a fixed-order DOB is proposed to meet model matching and disturbance rejection performance criteria. The estimated models of the multi-variable control system are employed to find the parameters of the compensator and filter.  The contributions of this work are summarized as follows:
\begin{itemize}
\item An iterative $H_\infty$ control algorithm for simultaneous design of the fully parameterized compensator and DOB is proposed in a convex optimization framework.
\item The infinite-norm constraints of the load current and input voltage disturbances sensitivity functions are considered in the constrained convex optimization problem.
\end{itemize}

The structure of the paper is as follows: Section II presents the control system of the closed-loop converter. Section III includes the formulation of the synthesis problem in a convex optimization framework. In Section IV, tracking performances of the controlled systems are presented under disturbance events and parameter changes of the converter. Section V presents the digital signal processor (DSP) based on real-time application results. Conclusions and discussions of the paper are given in Section VI.

\section{Low-frequency multi-variable models of the converter}
The power network of the IBC circuit is shown in Fig. 1, where $L_1$ and $L_2$ are the inductances with parasitic resistances $(r_l)$. $C_o$ is the filter capacitance with parasitic resistance ($r_c$) and $i_o(t)$ is the load current disturbance. $F_1$ and $F_2$ are the active switches of the relevant phase leg of the converter with on-state parasitic resistances ($r_s$). $E_1$ and $E_2$ are the passive switches of the relevant phase leg of the converter. $U$ is the DC duty cycle of the gate signals, $v_{in}(t)$ is the input voltage, $v_o(t)$ is the output voltage, $R_o$ is the load resistance and $i_{out}(t)$ is the load current of the converter. $V_{g1}(t)$ and $V_{g2}(t)$ are the gate signals of the relevant phase leg of the converter. The converter consists of four operating modes according to the on-off states of the active switches. The operating modes are presented in Fig. 1. The components of the converter are measured using an LCR meter to find the exact values of the components and parasitic resistances. The nominal and measured values of the converter are given in Table 1.  \par
 \begin{figure}[]
    \centering
    \includegraphics[width=8cm]{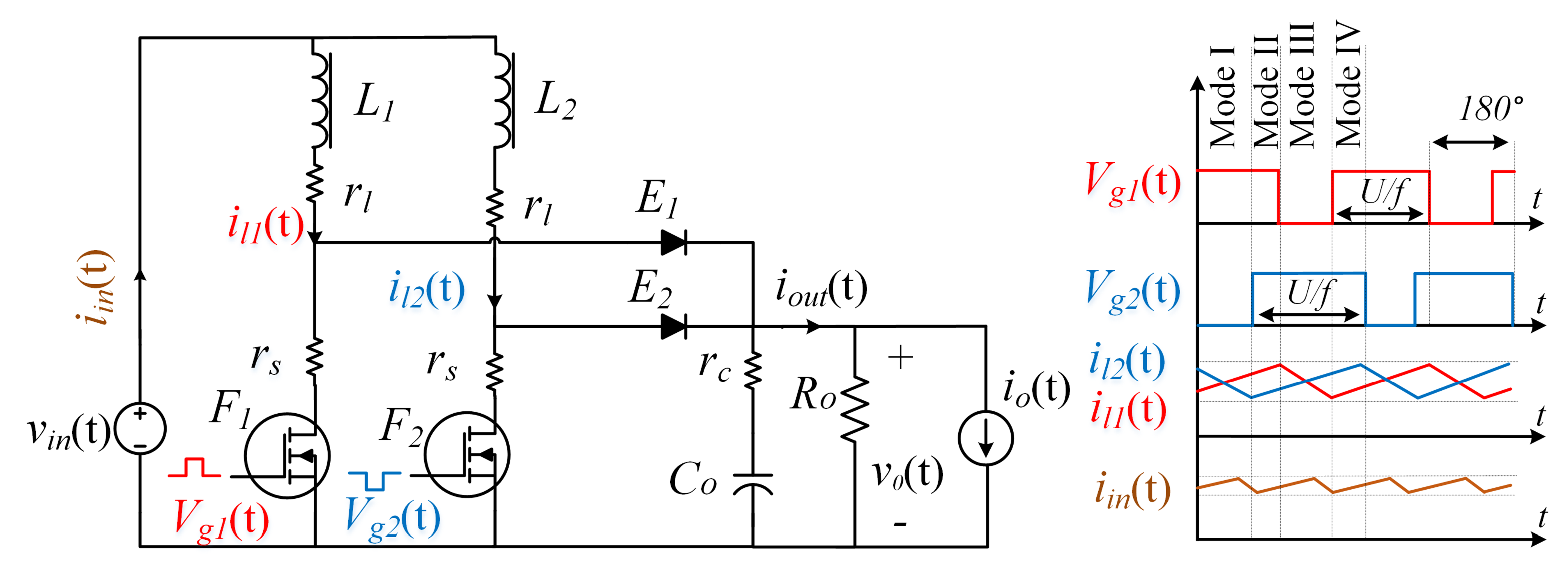}
    \caption{The power network of the interleaved boost converter for the duty cycle greater than $0.5$  }
    \label{fig1}
\end{figure}
 \begin{table}
\caption{Previous and updated system parameters of the 200 W prototype converter}
    \begin{tabular}{p{1cm}p{2cm}p{2cm}p{1cm}}
\centering
 Expr. & Pre. value & New value & Unit \\
 \hline \hline
 $L_{1}$   & 5 & 5.069  & mH  \\
 $L_{2}$   &  5 & 5.085  & mH  \\
 $C_{o}$ &  1  & 0.996 & mF  \\
 $r_{l}$ & 0.5  & 0.585  & \si{\ohm}  \\
$r_{s}$ & 0.036  & 0.036  & \si{\ohm} \\
$r_{c}$ & 0.05  & 0.01  & \si{\ohm} \\
$R_o$ & 50   & & \si{\ohm} \\
$v_{in}(t)$ & 46  & - & V \\
$v_{o}(t)$ & 100  & - & V \\
$f$ & 10 & - & kHz
    \end{tabular}
    \label{tab2}
\end{table}
The averaged non-linear models of the converter were derived according to the operating modes of the converter. The linear transfer functions of the multi-variable control system are obtained using the state-space averaging approach \cite{keskin2021multi}. The low-frequency transfer functions of the converter are given as
\begin{figure*}
\begin{align}
& G_{m}(s)=\frac{(-V_o(C_or_cs+1)(R_o^2(4U-2U^2-2)+R_o(r_c+r_{s})+r_c(r_l+r_{s})+LsR_{tot})}{(U-1)(C_oLs^2(R_o^2+2R_or_c+r_c^2)+C_oUR_os(R_o(r_{s}-r_c)-r_c^2+2r_{s}r_c))+T_{c} +T_{s}},\\
& G_{vm}(s)=\frac{-(2R_o(R_{tot})(U-1)(C_or_cs+1))}{C_oLs^2(R_o^2+2r_c+r_c^2)+C_oUR_os(R_o(r_{s}-r_c)+r_c(2r_{s}-r_c))+T_{c}+T_{s}}, \\
& G_{im}(s)=\frac{(R_{tot})(C_oR_os+C_or_cs+1)}{C_oLs^2(R_o^2+2r_c+r_c^2)+C_oUR_os(R_o(r_{s}-r_c)+r_c(2r_{s}-r_c))+T_{c}}+T_{s}, \\
& T_{c}=s(C_oR_o(R_o(r_c+r_l)+2r_cr_l+r_c^2)+C_or_c^2(Ur_{s}+r_l)+R_{tot}L),\\
& T_{s}=R_o^2U(2U-4)+UR_o(r_{s}-r_c)+Ur_{s}r_c+R_o(2R_o+r_c)+R_{tot}r_c, \\
& R_{tot}=R_o+r_c,  R_{oc}=R_o/R_{tot}.
\end{align}
\hrulefill
\end{figure*}

The DC magnitudes and AC deviations are denoted by capital and small letters with a tilde $(\sim)$, respectively. $\widetilde{u}(s)$ is the AC deviation of the duty cycle, $\widetilde{i}_o(s)$ is the AC deviation of the load current disturbance, $\widetilde{v}_{o}(s)$ is the AC deviation of the output voltage and $\widetilde{v}_{in}(s)$ is the input voltage disturbance. The mathematical transfer function of the plant ($G_{m}(s)$) can not be inverted since it has the right half plane (RHP) zero. Therefore, we use linear parametric system identification methods to obtain the minimum phase transfer function of the plant. Orthogonal pseudo-random binary sequences (PRBS) signals, which have a length of 1.000.000 samples and two times higher bandwidth than RHP zero, are applied sequentially to the open-loop converter as input voltage and gate signals. The PRBS signal applied to the active switches of the converter is presented in Fig. 2. The autoregressive moving average with exogenous inputs (ARMAX) approach is preferred to estimate the frequency responses of the converter. The estimation process is performed using the MATLAB identification toolbox. The order of the estimated plant model is equal to the mathematical model of the converter. The transfer functions of the multi-variable control system are given as

\begin{equation}
\small
\begin{split}
    & G_n(s)=\frac{-74.79s+1.811 \cdot 10^{7}}{s^2+284s+9.129\cdot10^{4}},\\
    & G_{i}(s)=\frac{-s-200}{0.001s^2+0.22s+104.6}, \\
    & G_{v}(s)=\frac{8.58s+1.724\cdot10^{5}}{s^2+277.8s+9.126\cdot10^{4}}, \\
    & W_n(s)=\frac{s^2-5730s+1.177\cdot10^{5}}{s^2+9600s+9.379\cdot10^{7}},
\end{split}
\end{equation} 
where $G_n(s)$ is the estimated transfer function of the plant from $\widetilde{u}(s)$ to $\widetilde{v_o}(s)$, $G_{i}(s)$ is the estimated transfer function of the load current disturbance, i.e from $\widetilde{i}_o(s)$ to $\widetilde{v}_o(s)$, $G_{v}(s)$ is the estimated transfer function of the input voltage disturbance, i.e from $\widetilde{v}_{in}(s)$ to $\widetilde{v}_o(s)$ and $W_n(s)$ is the transfer function of the output voltage sensor noise. The frequency responses of the mathematical and estimated transfer functions are presented in Fig. 3. \par
\begin{figure}
    \centering
    \includegraphics[width=7cm]{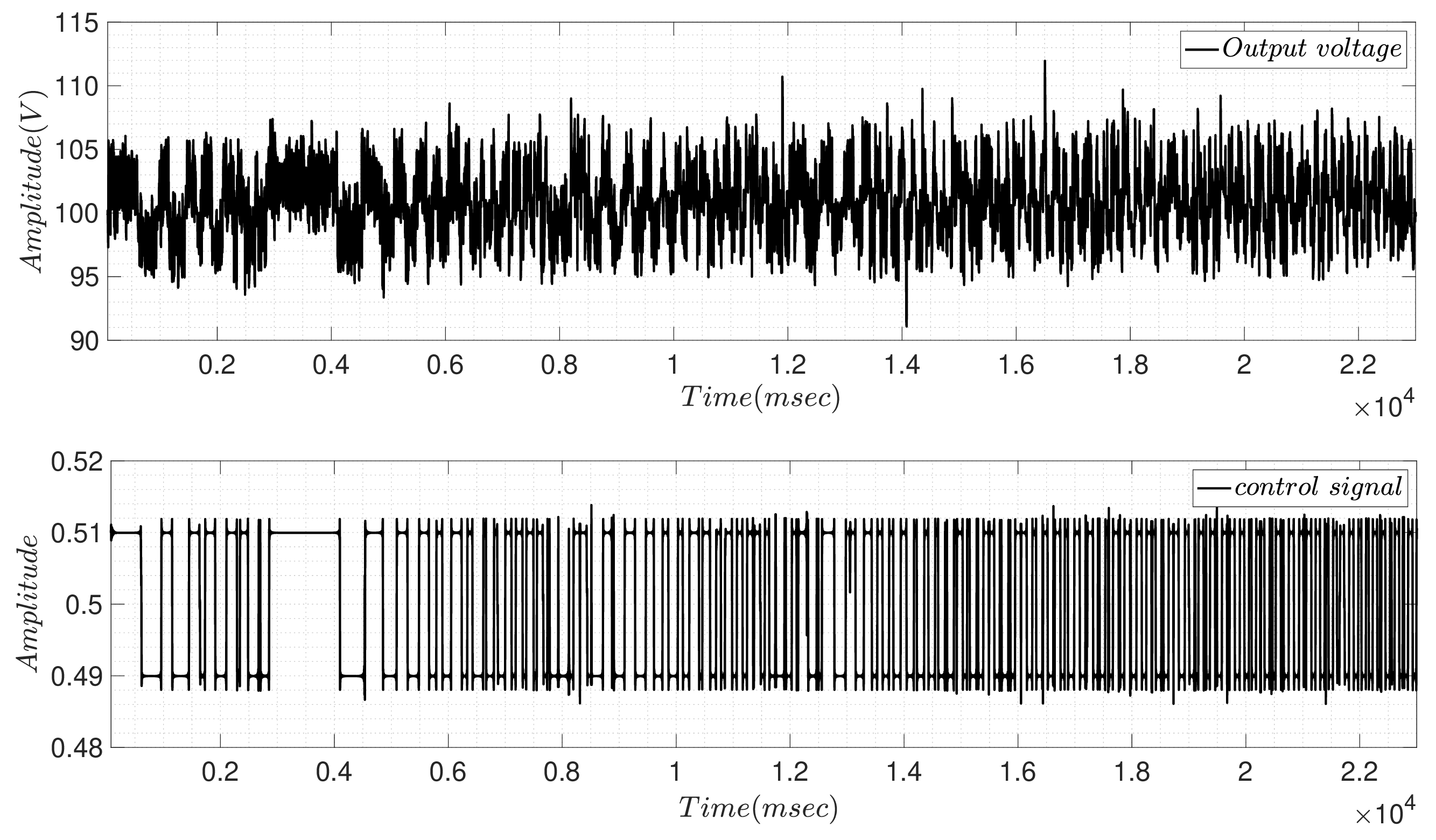}
    \caption{The time domain responses of the output voltage and PRBS signal}
    \label{fig4}
\end{figure}
\begin{figure}
    \centering
    \includegraphics[width=7cm]{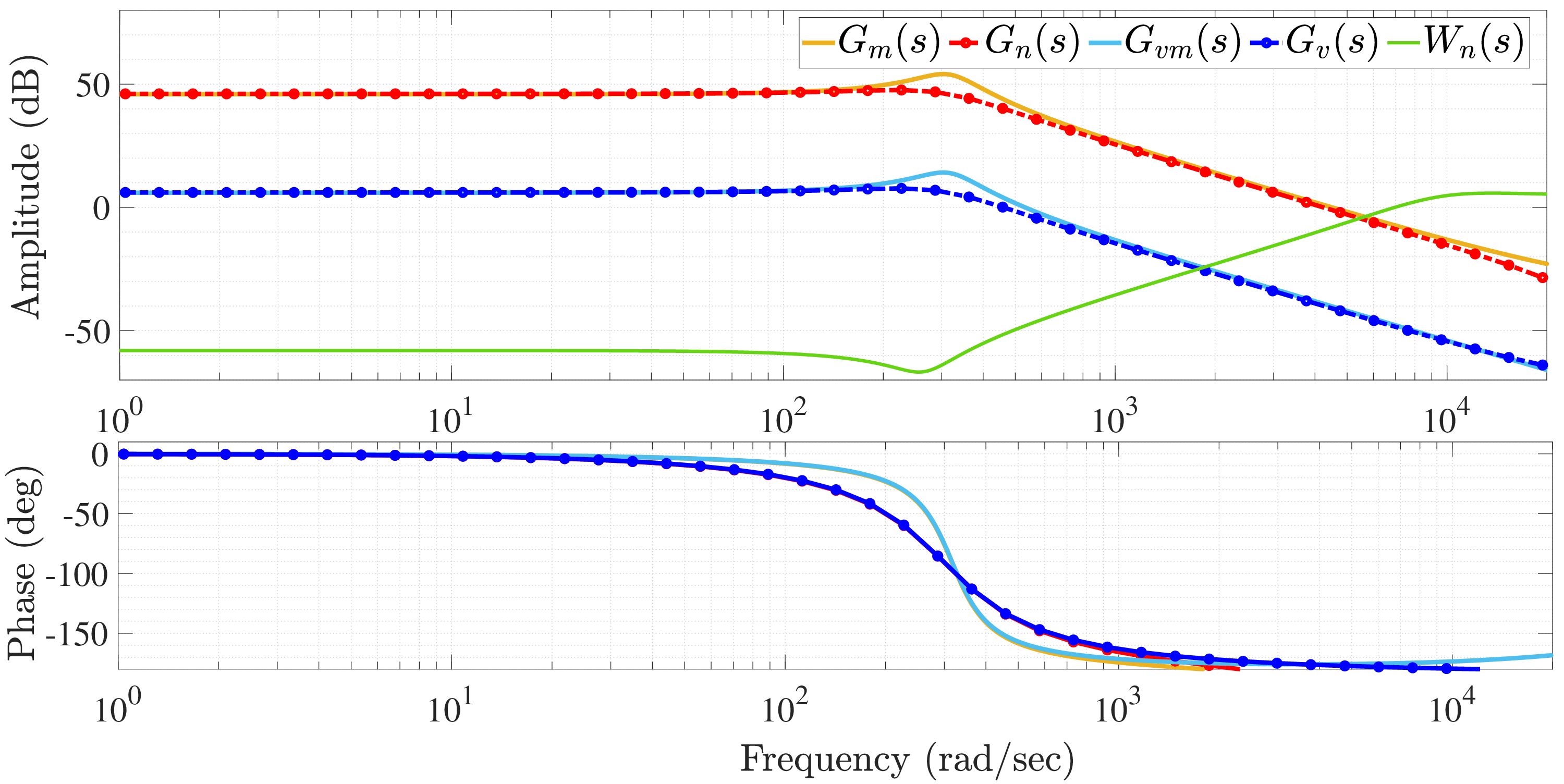}
    \caption{The frequency responses of the mathematical and estimated transfer functions}
    \label{fig2}
\end{figure}

The conventional 1-DOF control scheme is presented in Fig. 4 (a), where $K(s)$ is the feedback compensator, $G(s)$ is the uncertain plant model, $W_i(s)$ is the weighting function of the load current disturbance, $W_v(s)$ is the weighting function of the input voltage disturbance, $n$ is the noise, $r$ is the set-point, $e$ is the error, $u_f$ is the control input of the feedback loop, $y_m$ is the measured output variable, and $y_n$ is the noisy measured output variable. The feedback compensator is designed as a Type-III compensator which has two real zeros, poles, and pure integral action since it has inherent phase boost properties \cite{chan2015generalized}. The bandwidth of the 1-DOF control system is restricted since the plant has the RHP zero. Therefore, the disturbance rejection performance of the system is limited. \par
A two-degrees of freedom (2-DOF) control system with a DOB is preferred to improve the disturbance rejection capability and robustness of the system. The DOB control scheme is presented in Fig. 4 (b), where $u$ is the equivalent control input to the plant, $G_n^{-1}(s)$ is the inverse of the nominal plant model, $\hat{d}$ is the estimation of the disturbances and $Q(s)$ is the low-pass filter of the DOB. The filter is utilized to suppress noise at high frequencies and disturbances at low frequencies. The uncertain plant model can be equal to the nominal plant model ($G(s)=G_n(s)$) since the plant model is estimated experimentally. The mentioned two conventional control schemes are combined in a robust $H_\infty$ framework to include the transient response and disturbance rejection performance conditions. The robust $H_\infty$ control design scheme of the proposed approach is presented in Fig. 5, where $W_{1}(s)$ is the error weighting function, $W_2(s)$ is the feedback control input weighting function and $W_3(s)$ is the output variable weighting function. $z_1$ is the error performance output, $z_2$ is the performance output of the control input  of the feedback compensator, and $z_3$ is the performance output of the measured variable.
\begin{figure}
\centering
\subfloat[The conventional closed-loop voltage-mode 1-DOF control scheme]
{\includegraphics[width=8cm]{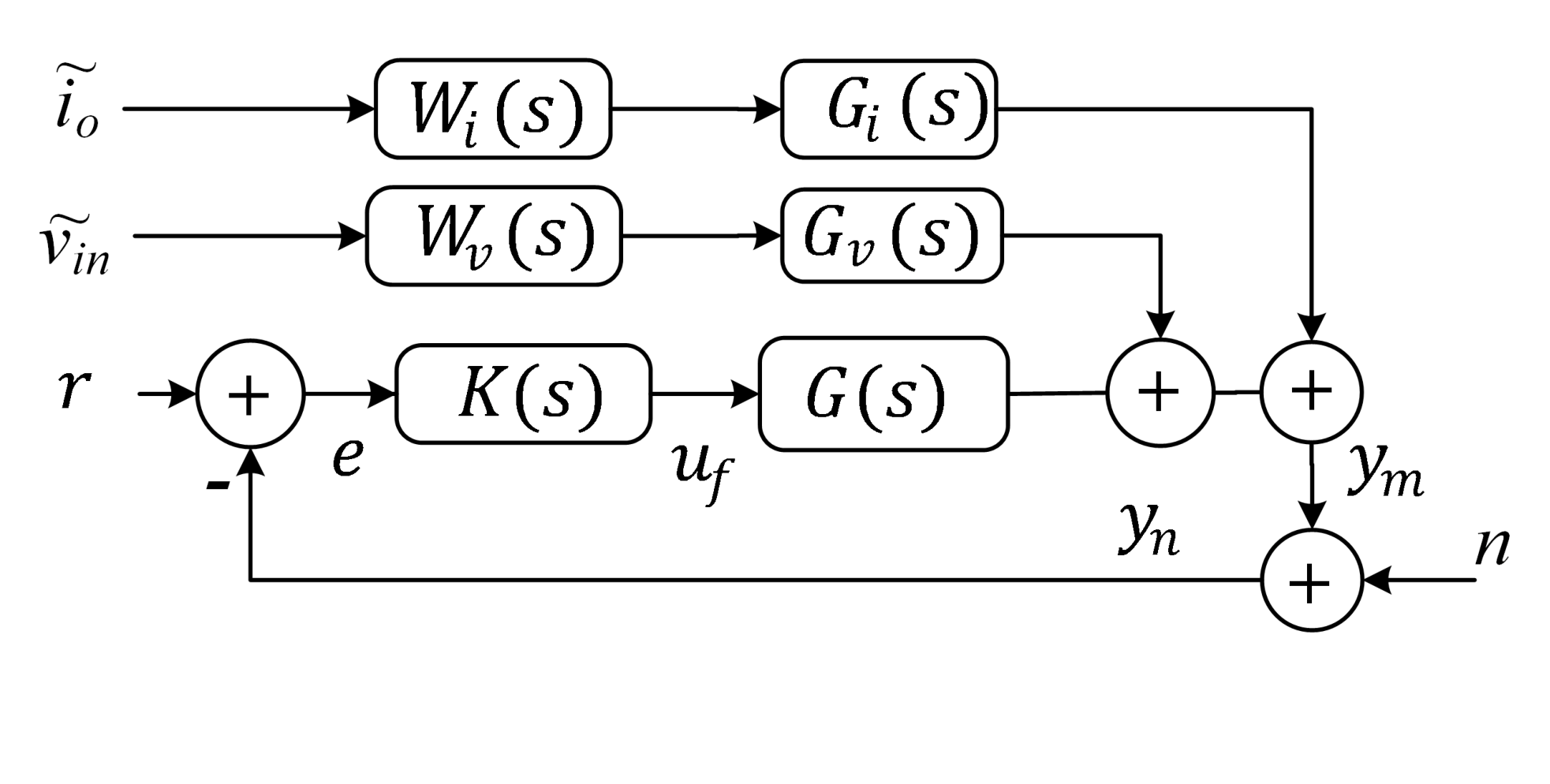}
\label{ber_vs_PS}} \\
\subfloat[The proposed DOB based 2-DOF control scheme]{\includegraphics[width=9cm]{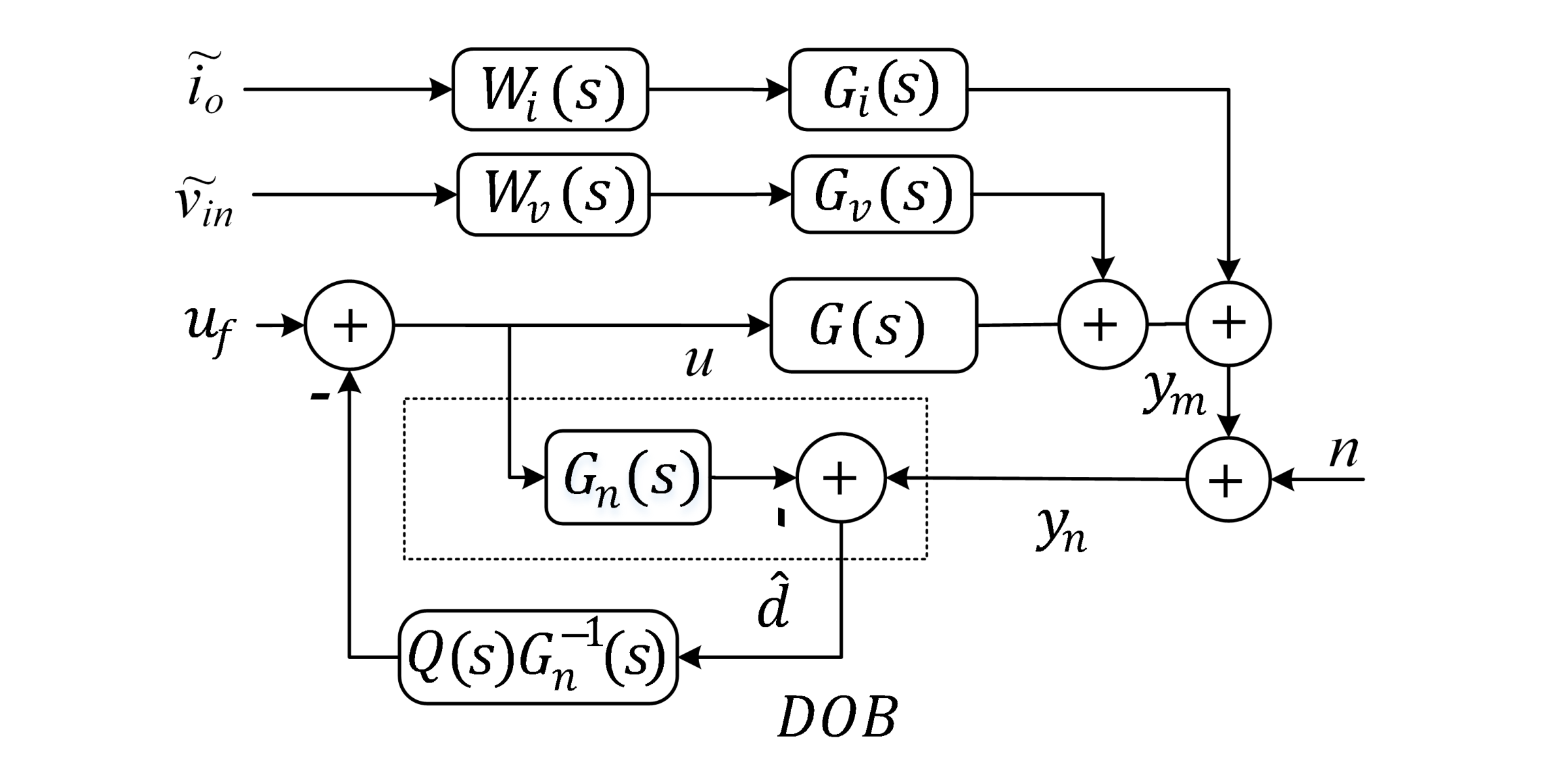}
\label{ber_vs_TS}} 
\caption{The conventional feedback and DOB control systems for the output disturbances}
\label{fig4}
\end{figure}
Note that the $G(s)$ is used instead of the nominal model transfer function for simplicity throughout the rest of the paper. The sensitivity functions of the outer feedback loop are obtained to structure the transient response ignoring the DOB loop \cite{keskin2023robust}. The three sensitivity functions are derived from Fig. 5 as
\\
\begin{equation}
    \begin{split}
       & S(s)=\frac{1}{1+G(s)K(s)}, \\
       & T(s)=\frac{-W_n(s)G(s)K(s)}{1+G(s)K(s)},\\
       & S_{i}(s)=\frac{K(s) }{1+G(s)K(s)},
    \end{split}
\end{equation}
where the sensitivity function $S(s)$ is the transfer function from set-point to error, i.e., $r$ to $e$, $T(s)$ is the complementary sensitivity function, which represents the transfer function from $n$ to $y_m$, and $S_{i}(s)$ is the input sensitivity function which represents the transfer function from $r$ to $u$. The sensitivity functions of the disturbance loops should be included in the problem to decrease singular values of the sensitivity function in the medium frequency area. Therefore, we can mitigate vital resonance modes of the control system. It enables the reduction of the output voltage oscillations and overshoots in the presence of uncertainties in the real parameters of the converter. The function of the output disturbances is derived as

\begin{equation}
S_{o}(s)=\frac{W_{v}(s)G_{v}(s)+W_{i}(s)G_{i}(s)}{1+G(s)K(s)},
\end{equation}
where $S_{o}(s)$ is the sensitivity function from $\widetilde{v}_{in}$ and $\widetilde{i}_{o}$ to ${y}_{m}$. It emphasizes that the parameter uncertainties in the plant cause the worst-gain effect at medium frequency area. Since disturbance transfer functions have resonance gain in the same frequency area, adding the functions in the problem could represent an additive uncertainty weighting function of the system. The feedback compensator is designed to obtain reference tracking and noise suppression performances. The inner DOB loop is included for further disturbance rejection and uncertainty suppression. The sensitivity functions of the inner DOB loop are given as

 \begin{equation}
\begin{split}
       & S_{D}(s)=\frac{W_i(s)G_i(s)+W_v(s)G_v(s)}{1+Q(s)},\\
       & T_{D}(s)=\frac{G(s)Q(s)W_n(s)}{1+Q(s)},
\end{split}
\label{e3}
\end{equation}
where $S_D(s)$ is the transfer function of the inner loop from $\widetilde{v}_{in}$ and $\widetilde{i}_{o}$ to $y_m$ and $T_D(s)$ is the transfer function of the inner loop from $n$ to $y_m$. These functions are utilized to improve the stability of the filter. The sensitivity functions, however, may be insufficient to shape the frequency response of the $Q(s)$-filter. The $Q(s)$ filter is shaped in the frequency domain using infinite norm constraint of the function $(W_Q(s)(1-Q(s))$ where $W_Q(s)$ reflects the frequency characteristics of the mismatched disturbances. It is highlighted that the function is a part of the numerator matrix of the closed-loop sensitivity function. The frequency characteristics of the function define the property of disturbance estimation \cite{wang2004design}. Assuming the multi-variable system is internally stable, DOB is employed to completely estimate and reject the disturbance inputs in the case $\lvert Q(j\omega) \rvert \approx 1$ at low and medium frequencies. 
\begin{figure}
    \centering
    \includegraphics[width=8cm]{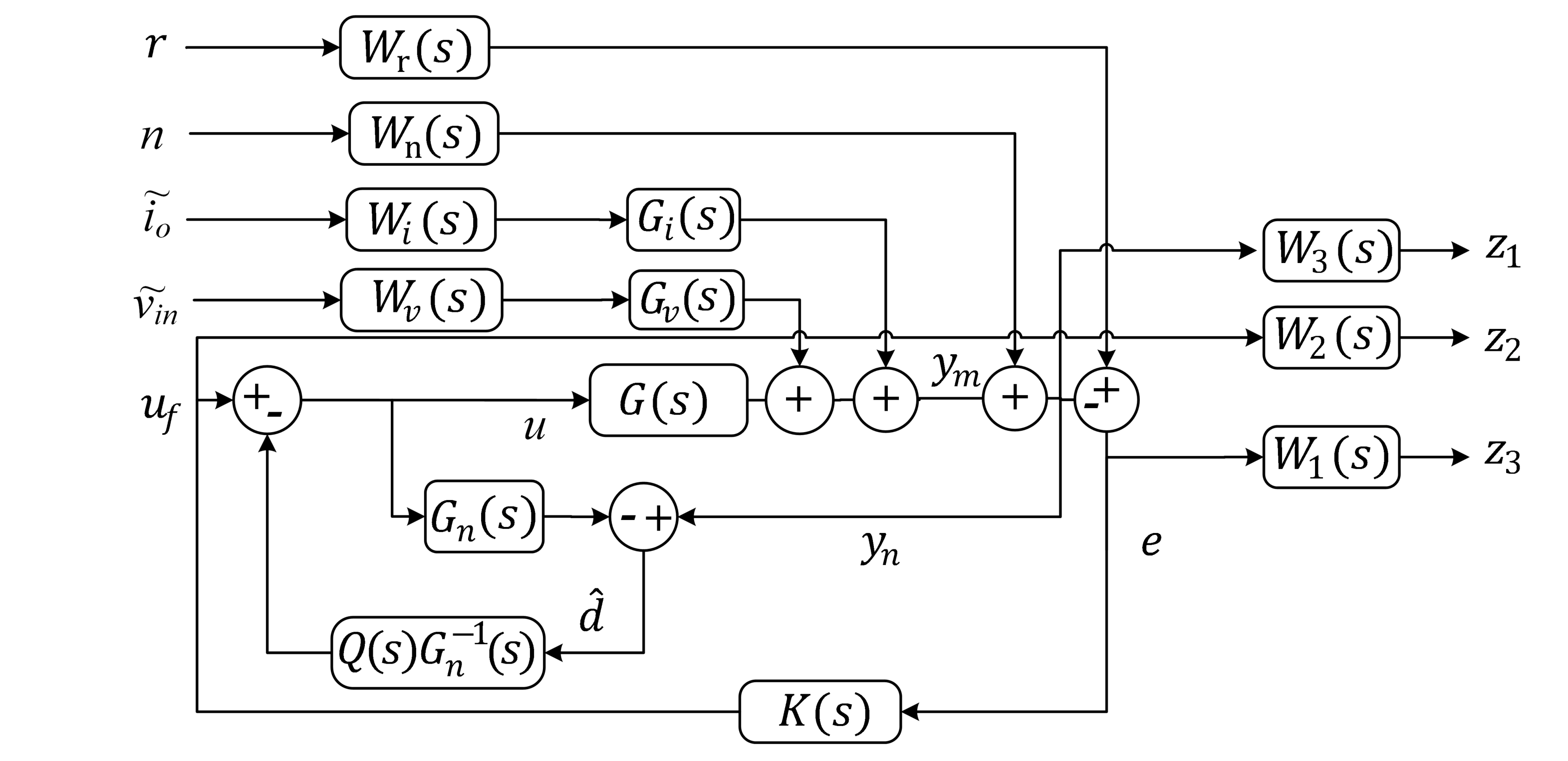}
    \caption{Re-structured control scheme for robust $H_\infty$ Type-III feedback compensator and DOB design problem}
    \label{fig4}
\end{figure}
The general sensitivity functions of the closed-loop system are given as

\begin{equation}
    \begin{split}
       & S_{C}(s)=\frac{(1-Q(s))(W_i(s)G_i(s)+W_v(s)G_v(s))}{1+G(s)K(s)},\\
       & T_{C}(s)=\frac{(Q(s)+G(s)K(s))W_n(s)}{1+G(s)K(s)},
    \end{split}
\end{equation}
where the sensitivity function $S_C(s)$ is the transfer function from the output disturbances to the measured variable, and the complementary sensitivity function $T_C(s)$ is the transfer function from the noise to the measured variable. For the stability of the inner and outer loop, the infinity norms of these two functions are also restricted.

\section{The convex optimization problem structure}
This section covers the convex optimization problem which is formulated to the simultaneous design of the feedback compensator and $Q(s)$ filter in the fixed-order $H_\infty$ framework. The control problem can be structured using the infinity norm representations of the mentioned sensitivity functions to satisfy the stability of the control system over defined disturbance effects. We now omit the complex frequency $(s)$ from this point on unless otherwise noted. The non-convex optimization problem is formulated as

\begin{equation}
\begin{array}{clclcl}
\displaystyle \min_{
\begin{matrix}
X, Y, N, M, \gamma \\
\end{matrix}} &\gamma, \\ [6mm]
\textrm{s.t.} &\big\|W_1S\big\|_\infty \leq \gamma,\\ [2mm]
&\big\|W_3T\big\|_\infty \leq \gamma,\\ [2mm]
&\big\|W_2S_i\big\|_\infty \leq \gamma,\\ [2mm]
&\big\|S_{o}\big\|_\infty \leq \gamma,\\ [2mm]
&\big\|S_{D}\big\|_\infty \leq \gamma,\\ [2mm]
&\big\|T_{D}\big\|_\infty \leq \gamma,\\ [2mm]
&\big\|W_{1C}S_{C}\big\|_\infty \leq \gamma,\\ [2mm]
&\big\|W_{3C}T_{C}\big\|_\infty \leq \gamma,\\ [2mm]
&\big\|(W_Q(1-Q))\big\|_\infty\leq \gamma, 
\end{array}
\end{equation}where $\gamma \in \mathbb{R}^{1\times1}$ is an auxiliary variable that represents the upper bounds of the singular values, $X$ is the numerator vector of the compensator, $Y$ is the denominator vector of the compensator, $N$ is the numerator vector of the filter, and $M$ is the denominator vector of the filter. Here, the integral action of the compensator can be achieved in two different ways. The integral action is added to the denominator of the compensator. Otherwise, the compensator can be simplified by adding the integral action to the nominal model. We prefer the second way to reduce the instability and numerical problems that may occur at certain frequency values in the compensator synthesis. Therefore, the matrices of the compensator and filter are expressed as

\begin{equation}
\begin{split}
&X(s)=[s^0 \ s^1 \ ... \ s^{h-1} \ s^{h}] \cdot [X_0 \ X_1 \ ... \ X_{h-1} \ X_{h}]^T, \\
&Y(s)=[s^0 \ s^1 \ ... \ s^{o-1} \ s^{o}] \cdot [Y_0 \ Y_1 \ ... \ Y_{o-1} \ 1]^T,\\
&N(s)=[s^0 \ s^1 \ ... \ s^{n-1} \ s^{n}] \cdot [N_{0} \ N_{1} \ ... \ N_{n-1} \ N_{n}]^T, 
 \\
&M(s)=[s^0 \ s^1 \ ... \ s^{m-1} \ s^{m}] \cdot [M_{0} \ M_{1} \ ... \ M_{m-1} \ M_{m}]^T,\\
\end{split}
\end{equation}
where $X \in \mathbb{R}^{h\times1}$, $Y \in \mathbb{R}^{o\times1}$, $N \in \mathbb{R}^{1\times n}$, and $M \in \mathbb{R}^{1\times m}$ for $h,\ o, \ n, \ m $ are the orders of the related coefficient vectors. The polynomial representations of the second-order feedback controller and first-order filter are given as
\begin{equation}
 \begin{split}
& K(s)= \frac{X_2s^2+X_1s+X_0}{Y_2s^2+Y_1s+Y_0}, \\
& Q(s)= \frac{N_0}{M_1s+M_0}.
\end{split}
\end{equation}
Equation (12) is a non-convex problem since it includes multiplications of compensator and filter matrices. These performance constraints can be transformed into iterative LMIs with the transition from the infinity-norm to second-norm representations. The problem is converted into LMIs in two separate stages, where the first stage is convex approximations of the feedback constraints and the second stage is the convex approximation of the DOB constraints. The sensitivity functions of the general system are used in both stages.

\subsection{The formulation of feedback loop constraints}
Consider the upper bounds of the sensitivity function with the infinity-norm that is given by
\begin{equation}
\big\|W_1S\big\|_\infty \triangleq \max_{\omega}\big(W_1S(j\omega)\big),
\end{equation}
where $max$ denotes the maximum singular value of the function. Therefore, the infinity-norm constraint can be written using $S=(1+GK)^{-1}$ as 
\begin{equation}
\footnotesize
\big\|W_1S\big\| \leq \gamma \Leftrightarrow \big(W_1(1+GK)^{-1}\big)^*(W_1(1+GK)^{-1}) \leq \gamma^2=\gamma_1, \\
\end{equation}
where $(\cdot)^*$ is the complex conjugate transpose and $\gamma_1$ is a non-negative real number. Equation (12) is rewritten using $K=XY^{-1}$ as
\begin{equation}
\begin{split}
& \big(W_1(1+GXY^{-1})^{-1}\big)^*\big(W_1(1+GXY^{-1})^{-1}\big) \leq \gamma_1, \\
& \big(YW_1\big)^*\big(YW_1\big) \leq \big(Y+GX\big)^*\gamma_1\big(Y+GX\big), \\
& (YW_1)^*(W_1Y)\underbrace{-(\underbrace{Y+GX}_\text{J})^*\gamma_1(Y+GX)}_\text{concave term} \leq 0.\\
\end{split}
\end{equation}
We still have quadratic matrix inequality due to the multiplication of compensator matrices. Therefore, we define a new variable $J=Y+GX$ as
\begin{equation}
\big(YW_1\big)^*(W_1Y)-J^*J\gamma_1 \leq 0. 
\end{equation}
This inequality is a quadratic convex-concave form. It can be linearized using first-order Taylor expansion of $J^{*}J$ around a feasible initial point. This equation corresponds to the factorization of a mathematical quadratic function. The $J^*J$ vector polynomial is equivalent to:
\begin{equation}
 J^*J\approx J^*J_i+J_i^*J-J_i^*J_i, 
\end{equation}
where the left-hand side polynomial is a quadratic function and the right-hand side term is a linear function around $J_i=Y_i+GX_i$ initial point. Multiplying (18) with $-\gamma_1^{-1}$ and combining with (19), the LMI representation of the constraint is given using the Schur complement lemma:
\begin{align}
\left[ {\begin{array}{cc} 
J^*J_i+J_i^*J-J_i^*J_i & (W_1Y)^*   \\
  W_1Y   &    \gamma_1   
\end{array} }
\right]\succeq 0.
\end{align}
The other constraints of the feedback loop are convexified using the steps given between (15-21). Finally, The LMI constraints for the optimization of feedback controller are given as 

\begin{eqnarray}
\begin{aligned}
& \left[ {\begin{array}{cc}
J^*J_i+J_i^*J-J_i^*J_i & (W_1Y)^*   \\
   W_1Y   &    \gamma_1    \\
\end{array}}
\right]\succeq 0, \\ 
& \left[ {\begin{array}{cc}
J^*J_i+J_i^*J-J_i^*J_i & (W_3GX)^*   \\
   W_3GX   &    \gamma_1    \\
\end{array}}
\right]\succeq 0, \\
& \left[ {\begin{array}{cc}
J^*J_i+J_i^*J-J_i^*J_i & (W_2X)^*   \\
   W_2X   &    \gamma_1    \\
\end{array} }
\right]\succeq 0,\\
& \left[ {\begin{array}{cc}
J^*J_i+J_i^*J-J_i^*J_i & \Big((W_{v}G_{v}+W_{i}G_{i})Y\Big)^*   \\
   (W_{v}G_{v}+W_{i}G_{i})Y   &    \gamma_1    \\
\end{array}}
\right]\succeq 0. 
    \end{aligned}
\end{eqnarray}

\subsection{The derivation of DOB constraints}
The same analysis can be applied to (12) for the sensitivity and complementary sensitivity functions of the DOB loop. For the sensitivity function of the DOB, the infinite norm constraint is given as
\begin{equation}
\big\|S_D\big\|_\infty \triangleq \max_{\omega}\big(S_D(j\omega)\big).
\end{equation}
By employing the same steps given between (15-21), the following equations are obtained
\begin{equation}
\footnotesize
\big\|S_D\big\| \leq \gamma_D \Leftrightarrow \Bigg(\frac{(W_iG_i+W_vG_v)M}{N+M}\Bigg)^*\Bigg(\frac{(W_iG_i+W_vG_v)M}{N+M}\Bigg)\leq \gamma_D^2 =\gamma_2,
\end{equation}
\begin{equation}
\footnotesize
((W_iG_i+W_vG_v)M)^*((W_iG_i+W_vG_v)M)-(N+M)^*\gamma_2(N+M)\leq 0,
\end{equation}
\begin{equation}
\footnotesize
\begin{split}
& (\underbrace{N+M}_\text{H})^*\gamma_2(N+M)-((W_iG_i+W_vG_v)M)^*((W_iG_i+W_vG_v)M) \geq 0, \\
& H^*H\approx H^*H_i+H_i^*H-H_i^*H_i, \\
\end{split}
\end{equation}
where $H_i=N_i+M_i$ is a initial point of $H$ variable. Then, using the Schur complement lemma, the following LMI representation is obtained
\begin{equation}
\left[{ \begin{array}{cc}
H^*H_i+H_i^*H-H_i^*H_i & ((W_iG_i+W_vG_v)M)^*   \\
   ((W_iG_i+W_vG_v)M)   &    \gamma_2    \\
\end{array}}
\right]\succeq 0.
\end{equation}The LMI representation of the constraint of the complementary sensitivity function, $T_C$, can be obtained following the steps given between (22-26). The LMI representations of the two sensitivity constraints are given as
\begin{equation}
\footnotesize
\left[ { \begin{array}{ccc}
H^*H_i+H_i^*H-H_i^*H_i & ((W_iG_i+W_vG_v)M)^* & (GW_nN)^*   \\
   ((W_iG_i+W_vG_v)M)   &    \gamma_2 &    0    \\
   GW_nN   &    0  &    \gamma_2   \\
\end{array}}
\right]\succeq 0.
\end{equation}
The convex constraints are utilized to improve the stability of the DOB loop. The following constraint is added to the problem to shape the filter. For the additional constraint of the DOB loop, the constraint is given as
\begin{equation}
\big\|W_Q(1-Q)\big\|_\infty \triangleq \max_{\omega}\big(W_Q(j\omega)(1-Q(j\omega))\big).
\end{equation}
By employing the same steps given between (16-19), the following equations are obtained as
\begin{equation}
\footnotesize
\big\|W_Q(1-Q)\big\| \leq \gamma_D \Leftrightarrow \big(W_Q(M-N)\big)^*W_Q(M-N)\leq (M^{-1})^*\gamma_2M^{-1},
\end{equation}
\begin{equation}
\begin{split}
& M^{*}\gamma_2M-\big(W_Q(M-N)\big)^*W_Q(M-N)\geq 0, \\
& M^*M \approx M^*M_i+M_i^*M-M_i^*M_i, \\
\end{split}
\end{equation} where $M_i$ is the initial point of the denominator vector of the filter. Using the Schur Complement lemma, the LMI representation of the constraint is obtained as
\begin{equation}
\left[ { \begin{array}{cc}
M^*M_i+M_i^*M-M_i^*M_i & \Big(W_Q(M-N)\Big)^*   \\
   W_Q(M-N)   &    \gamma_2    \\
\end{array} }
\right]\succeq 0.
\end{equation}

\subsection{The derivation of general loop constraints}
For the general stability of the control system, the infinity norm of sensitivity functions of the closed-loop system should be minimized. In this subsection, we convert the non-convex constraints of sensitivity functions into LMIs assuming that the matrices of the compensator are pre-computed. In other words, the matrices of the filter are considered fixed when the compensator matrices are considered variables or vice-versa. For the sensitivity of the closed loop, the constraint is given as
\begin{equation}
\big\|S_C\big\|_\infty \triangleq \max_{\omega}\big(W_{1C}S_C(j\omega)\big).
\end{equation}
By employing the steps given between (12-15), the following equations are obtained
\begin{equation}
\footnotesize
\big\|W_{1C}S_C\big\| \leq \gamma_C \Leftrightarrow \Big(\frac{P\big(1-\frac{N}{M}\big)}{1+\frac{GX}{Y}}\Big)^*\Big(\frac{P\big(1-\frac{N}{M}\big)}{1+\frac{GX}{Y}}\Big)\leq \gamma_C^2=\gamma_3,
\end{equation}
\begin{equation}
\big(PY(M-N)\big)^*\big(PY(M-N)\big)-(MJ)^*\gamma_3 MJ\leq 0,
\end{equation}
\begin{equation}
\begin{split}
& (MJ)^*\gamma_3 MJ-(PY(M-N))^*PY(M-N)\geq 0, \\
& M^*M \approx M^*M_i+M_i^*M-M_i^*M_i, \\
\end{split}
\end{equation}
where $W_{1C}(W_iG_i+W_vG_v)$ function is denoted by $P$ to simplify the equations. It is highlighted that only the $M^*M$ polynomials are linearized around $M_i$ initial point since assuming the compensator matrices are fixed. Then, using the Schur Complement lemma, the following LMI representation is obtained as
\begin{equation}
\small
\left[ { \begin{array}{cc}
J^*J(M^*M_i+M_i^*M-M_i^*M_i) & (PY(M-N))^*   \\
   PY(M-N)   &    \gamma_3    \\
\end{array} }
\right]\succeq 0.
\end{equation}
The LMI representation of the constraint of the complementary sensitivity function, $T_C$, can be obtained following the steps (28-32). Finally, convex approximations of the general closed-loop constraints are given as
\begin{equation}
\footnotesize
\left[ {\begin{array}{ccc}
J^*J(M^*M_i+M_i^*M-M_i^*M_i) & \big(PY(M-N)\big)^* & \big(W_T(MGX+NY)\big)^*   \\
PY(M-N)   &    \gamma_3 &    0    \\
   W_T(MGX+NY)   &    0  &    \gamma_3    \\
\end{array} }
\right]\succeq 0,
\end{equation}
where $W_T$ represents $W_{3C}W_n$ function. The structured CCP algorithm for the 2-DOF control system is proposed in Algorithm 1, where $N$ and $M$ matrices are equal to the $N_i$ and $M_i$ matrices only in the first iteration of stage 1. $\varepsilon$ is a pre-specified tolerance, $i$ is the iteration number, and $l$ is the maximum iteration number. We organize the proposed algorithm as two stages since the matrices $X$ and $M$ or the matrices $Y$ and $N$ are in multiplication. The $\gamma_3$ can be $\gamma_1$ or $\gamma_2$ according to the stage of the algorithm in which the compensator or filter is optimized. This problem is solved by employing a sequential solution in a generic loop. Finally, the coefficients of the compensator and filter are simultaneously optimized due to the proposed algorithm, which is solved in a similar way to the classical $D-K$ iteration method in robust control.
\begin{figure}
    \centering
    \includegraphics[width=8cm]{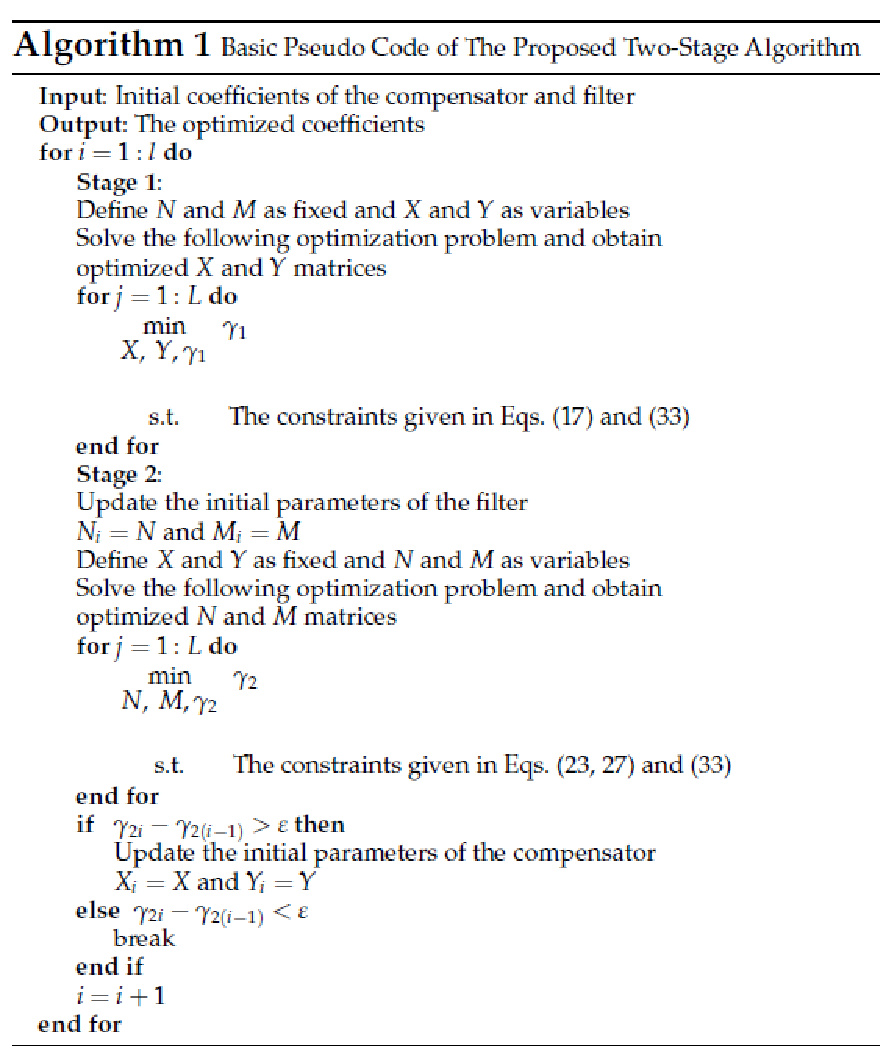}
\end{figure}

\subsection{K-factor method}
The K-factor method is preferred to design of a Type-III compensator the performances of the proposed 1-DOF and 2-DOF control systems. The method is the only analytical method used to synthesize the compensator. The method is based on the principle of placing the zero and poles of the compensator according to the designer' preference by analyzing the amplitude-frequency curve of the nominal system. It is determined the required phase margin and amplitude gain boost at a selected crossover frequency. The phase boost expression is given as 
\begin{equation}
 A_p=2(tan^{-1}(\frac{f_c}{f_{z_{1, 2}}})-tan^{-1}(\frac{f_c}{f_{p_{1, 2}}}))-\frac{\pi}{2},
\end{equation}
where $A_p$ is the desired phase boost value in degree, $f_c$ is the crossover frequency, $f_{z_{1, 2}}$ are the frequencies of the zeros, $f_{p1, 2}$ are the frequencies of the poles. The $k$ gain of the method is given as
\begin{equation}
k={(tan(\frac{ A_p}{4}+\frac{\pi}{4}))^2}.    
\end{equation}
The coincident zero pair of the compensator is given as
\begin{equation}
f_{z_{1, 2}}=\frac{f_c}{\sqrt{k}}=\frac{f_c}{tan(\frac{A_p}{4}+\frac{\pi}{4})}.    
\end{equation}
The coincident pole pair of the compensator is given as
\begin{equation}
f_{p_{1, 2}}=f_c \sqrt{k}=tan(\frac{ A_p}{4}+\frac{\pi}{4})f_c.    
\end{equation}
The mid-band gain of the compensator is derived as

\begin{equation}
G_o=\frac{\sqrt{1+(\frac{f_c}{f_{p_1}})^2}\sqrt{1+(\frac{f_c}{f_{p_2}})^2}}{\sqrt{1+(\frac{f_{z_1}}{f_c})^2}\sqrt{1+(\frac{f_c}{f_{z_2}})^2}} M_{g},   
\end{equation}
where $M_{g}$ is the gain value at the selected crossover frequency. Finally, the compensator is given as
\begin{equation}
K_{k}(s)=G_o\frac{(s/2\pi f_{z1}+1)(s/2\pi f_{z2}+1)}{(s/2\pi f_{z1}+1)(s/2\pi f_{z2}+1)}, 
\end{equation}
where $K_{k}(s)$ is the Type-III compensator designed by using K-factor method. \cite{rana2017development}.

\subsection{Implementation stage } 
The convex optimization problem is semi-infinite programming since the problem is defined in the infinite frequency range. By searching for a solution in a frequency range higher than the closed-loop bandwidth, it is transformed into a semi-definite programming problem, where any LMIs solver could be used for the solution of the problem. The problem is solved in a reasonable frequency number $(L)$ using an inner loop for frequency gridding. In other words, the problem is discretized with a logarithmically spaced frequency domain into a finite set $w_j = {w_1, \ w_2, \ ..., \ w_L}$ where $j$ is the number of the inner loop cycles. The reasonable gridding number could be found using the randomized scenario approach to ensure the constraints with sufficient probability level \cite{campi2018introduction, dacs2021robust}. The reasonable number for scenarios given as
\begin{equation}
L \geq \frac{2}{\epsilon_r}(d_p-1+\ln{\frac{1}{\beta_c}}),
\end{equation}
where $d_p$ is the number of optimization variables, $\epsilon_r \in (0,1)$ is the risk index, and $\beta_c \in (0,1)$ is the confidence index. The convex-concave procedure presents a robust solution to original non-convex optimization problems, although it does not guarantee convergence to the global solution of the non-convex problem. Robust stability and performance improvements are achieved by selecting the relevant weighting functions. In convex-concave programming, the design must start from feasible initial coefficients of compensator and filter which may be a challenge to find while the basic iteration will work in all of these variations. In addition, it is recommended that the initial coefficients of the compensator should be close to zero for a successful optimization process. In the selection of the coefficients, it should be noted that the zeros and poles of the compensator should be negative real values. Alternatively, the initial coefficients of the compensator can be found via the K-factor method which is a classical approach for the design of the compensator or any $H_\infty$ solver such as $hinfstruct$ \cite{karimi2017data, dacs2021data}. 

\section{Simulation results of the converter}
The two-phase converter circuit is designed for continuous current mode operation for medium and high power transmission conditions. The values of the passive components are preferred as high values so that the converter remains in continuous current mode over the entire load operating range. The computations of the constrained convex optimization problem are solved by the convex modeling framework YALMIP using the MOSEK software \cite{lofberg2008modeling}. The problem is solved with 200 grinding points in the $10^{2}-10^{5}$ rad/s frequency range. The initial coefficients of the compensator are chosen as 0.0001 so that the initial compensator is equal to an integrator. The initial coefficients of the filter are chosen as $N_i$=120 and $M_i$=[1 120]. The achieved objectives values are $\gamma_1=0.813$ and $\gamma_2=1$. In the K-factor method, high phase contribution is required since the phase margin and amplitude gain of the system are low based on the amplitude-frequency response of the open-loop converter which is presented in Fig. 3. The compensator is designed for a phase margin over 70 degrees and a gain margin over 12 dB, all of which are regarded as typical requirements for disturbance rejection performance of a converter system. Therefore, the oscillations of the output voltage during disturbance events can be mitigated. The compensator coefficients are obtained by assuming that the required phase boost ($A_p$) is 172 degrees and gain loss ($M_{g}$) would be -40 dB at 2300 Hz crossover frequency. \\
The aforementioned weighting functions are selected as follows:
\begin{equation}
\begin{split}
    & W_1(s)=\frac{s+395}{1.4s+3.95\cdot10^{-7}}, \ \ W_2(s)=\frac{s+1850}{0.04s+18500}, \\[2mm]
    & W_3(s)=\frac{s+400}{15s+2000}, \ \ W_{v}(s)=\frac{8\cdot10^{-8}s+3200}{s+2094}, \\[2mm]
    & W_{i}(s)=\frac{1\cdot10^{-8}s+400}{s+2094}, \ \ W_{Q}(s)=\frac{s+140}{1.4s+1.4}, \\
    & W_{1C}(s)=\frac{s+3700}{1.3s+0.0000271}, \ \ W_{3C}(s)=\frac{s+3700}{0.04s+18500}.
\end{split}
\end{equation}
In order to reduce the effect of converter parameter changes on the output voltage, the amplitude, and frequency values of the input voltage and load current weighting functions are selected high. Finally, the optimal coefficients of the compensators and the filter are given as
\begin{equation}
\begin{split}
        & K_{x}(s)=\frac{918.9s^2+3.558\cdot10^{5}s+3.439\cdot10^{7}}{s^3+1.37\cdot10^{4}s^2+4.687\cdot10^{7}s},\\[2mm]
        & K_{k}(s)=\frac{5.472\cdot10^{7}s^2+5.282\cdot10^{10}s+1.275\cdot10^{13}}{482.7s^3+3.822\cdot10^{8}s^2+7.564\cdot10^{13}s},\\
        & K_{2x}(s)=\frac{1692s^2+4.1\cdot10^{5}s+1.47\cdot10^{8}}{s^3+1.4\cdot10^{4}s^2+10^{8}s}, \\
        & Q(s)= \frac{98.77}{s+99.6},\\
\end{split}
\end{equation}
where $K_x$ is the compensator of the 1-DOF system and $K_{2x}$ is the compensator of the 2-DOF system which is optimized with the proposed method.
\begin{figure}
    \centering
    \includegraphics[width=8cm]{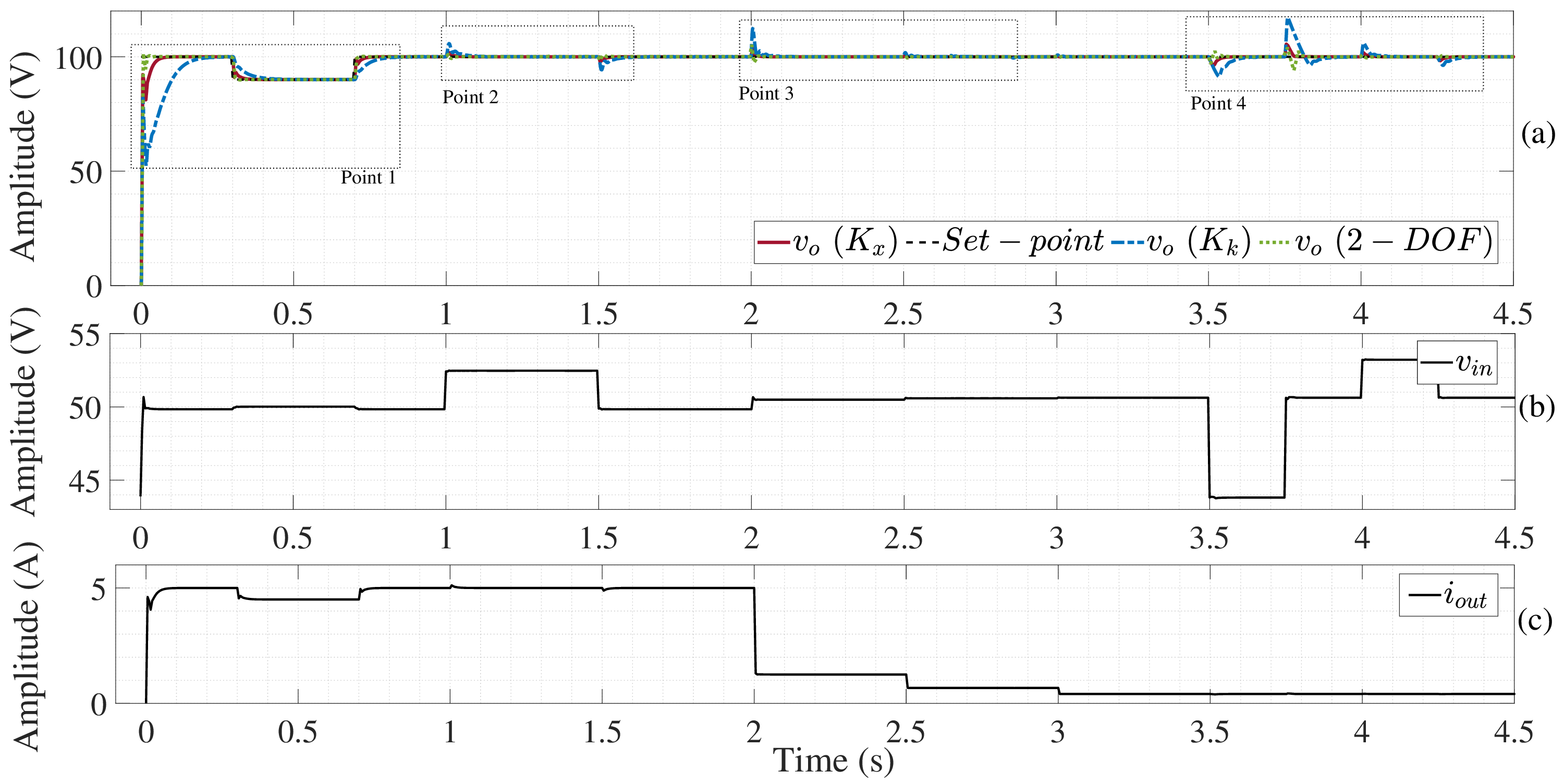}
    \caption{The performance results of the controlled systems in the time domain: (a) Set-point tracking performances (b) The input voltage changes  (c) The load current changes} 
    \label{fig6}
\end{figure}

The low-high power performances of the controlled closed loop converter are presented in Fig. 6. The control systems guarantee to track the variable set point. However, the control systems present different performances in problematic disturbance events. To examine the performances of the control systems in detail, Fig. 6 is divided into four different operating points as in Figs. 7 and 8. The set-point tracking performances of the control systems at starting of the converter are presented in Fig. 7. The settling time of the $K_{k}$ controlled system is three times higher than the settling time of the proposed 1-DOF control system. The set-point of the output voltage is changed with a range of 10$\%$ at the operation point stepwise from $v_o=100$ V to $v_o=90$ V at $t=0.3$ s and from $v_o=90$ V to $v_o=100$ V at $t=0.7$ s. The control systems have no overshoots at the starting and the changes of the set-point (Point 1). The input voltage disturbance rejection performances of the control systems are presented at high and low power transmissions in Figs. 7 and 8 (Points 2, 4). 
\begin{figure}
\centering
\includegraphics[width=8cm]{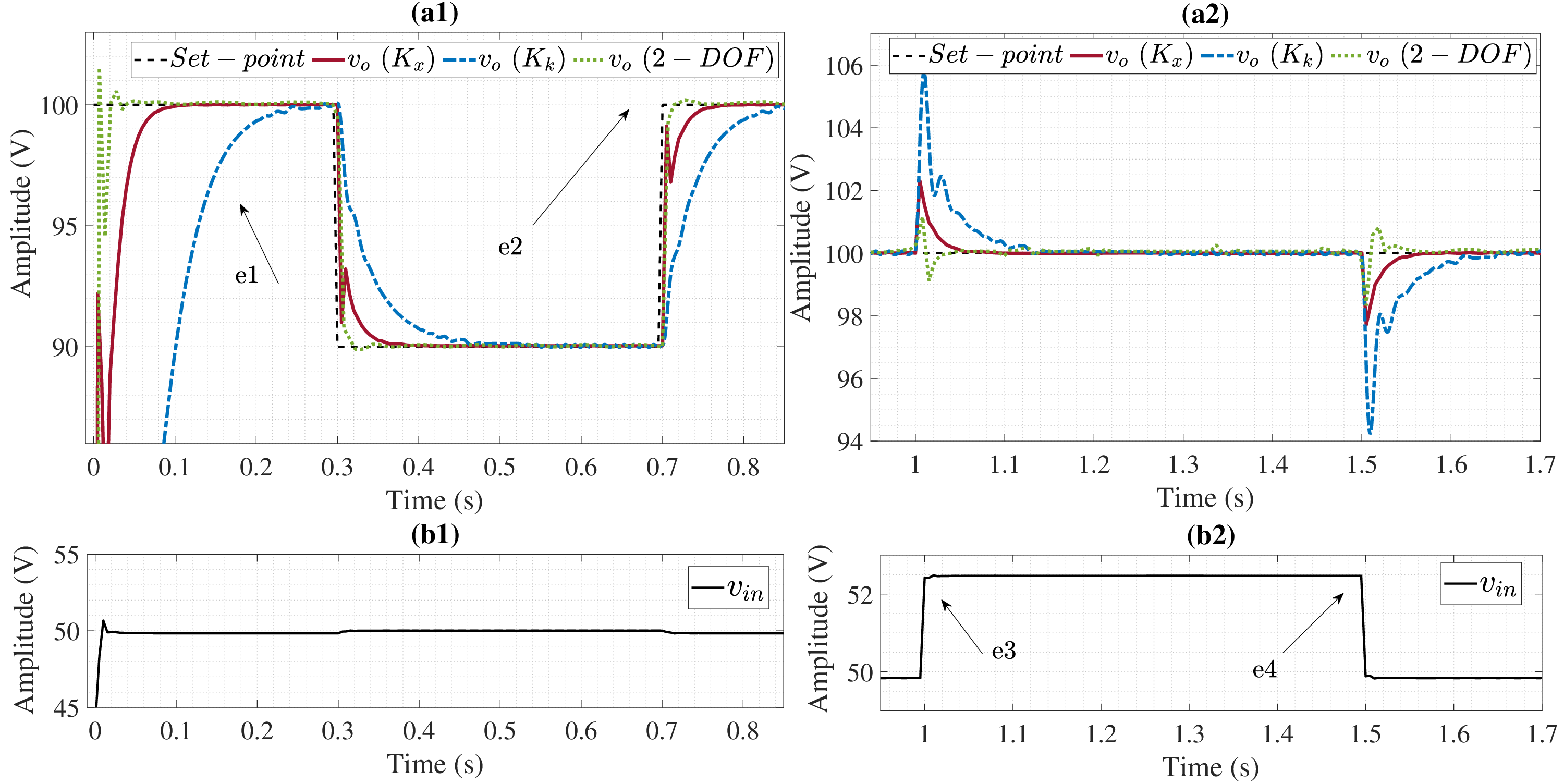}
\caption{The performance results of the control systems under the problematic events: (a1, b1) Point 1: Set-point tracking at the start-up and around the nominal operating point ($i_{out}= 5$ A), (a2, b2) Point 2: Step changing of input voltage at high power transmission ($i_{out}= 5$ A) } 
\label{fig6}
\end{figure}

Fig. 8 demonstrates the rejection performances of the load disturbance effects on the output voltage caused by varying the load current around high and low-power working conditions. The load resistance is changed stepwise from $R_o=20 \ \si{\ohm}$ to $R_o=80 \ \si{\ohm}$ at $t=2$ s, from $R_o=80 \ \si{\ohm}$ to $R_o=150 \ \si{\ohm}$ at $t= 2.5$ s, and from $R_o=150 \ \si{\ohm}$ to $R_o=240 \ \si{\ohm}$ at $t= 3$ s (events 5, 6 and 9). The proposed 1-DOF and 2-DOF control systems have nearly no oscillation under the external disturbance events due to boosting the phase margin of the controlled systems to the requested level. The $K_{x}$ controlled system has nearly 80 ms under set-point changes and 50 ms settling times under input voltage and load current changes. The $K_{k}$ controlled system, however, has 230 ms and 140 ms settling times under set-point input voltage, and load current  changes, respectively. The inductances and related parasitic resistances of the converter are changed stepwise from the nominal value to half of the nominal value at 2.65 s in Fig. 8 to show the uncertainty rejection performances of the controlled systems (event 7). At 2.85 s the components are increased stepwise to the nominal values (event 8). The effect of the events on the output voltage is ignored. The quantitative performance metrics, i.e. settling time and maximum overshoots, are introduced to assess the control performances. The performance metrics of the simulation results under set-point tracking and disturbance rejection are given in Table 2. It is seen that the proposed 1-DOF control system decreases the overshoot 60$\%$ and the settling time 81$\%$ according to the $K_{k}$ controlled system considering point 3. The 2-DOF control system presents solid uncertainty rejection performance in the presence of changes in converter parameters. The system has minimum settling times at the parameter changes. However, proposed 1-DOF and 2-DOF control systems seem to have similar disturbance rejection performances at the load current changes.

\begin{figure}
\centering
\includegraphics[width=8cm]{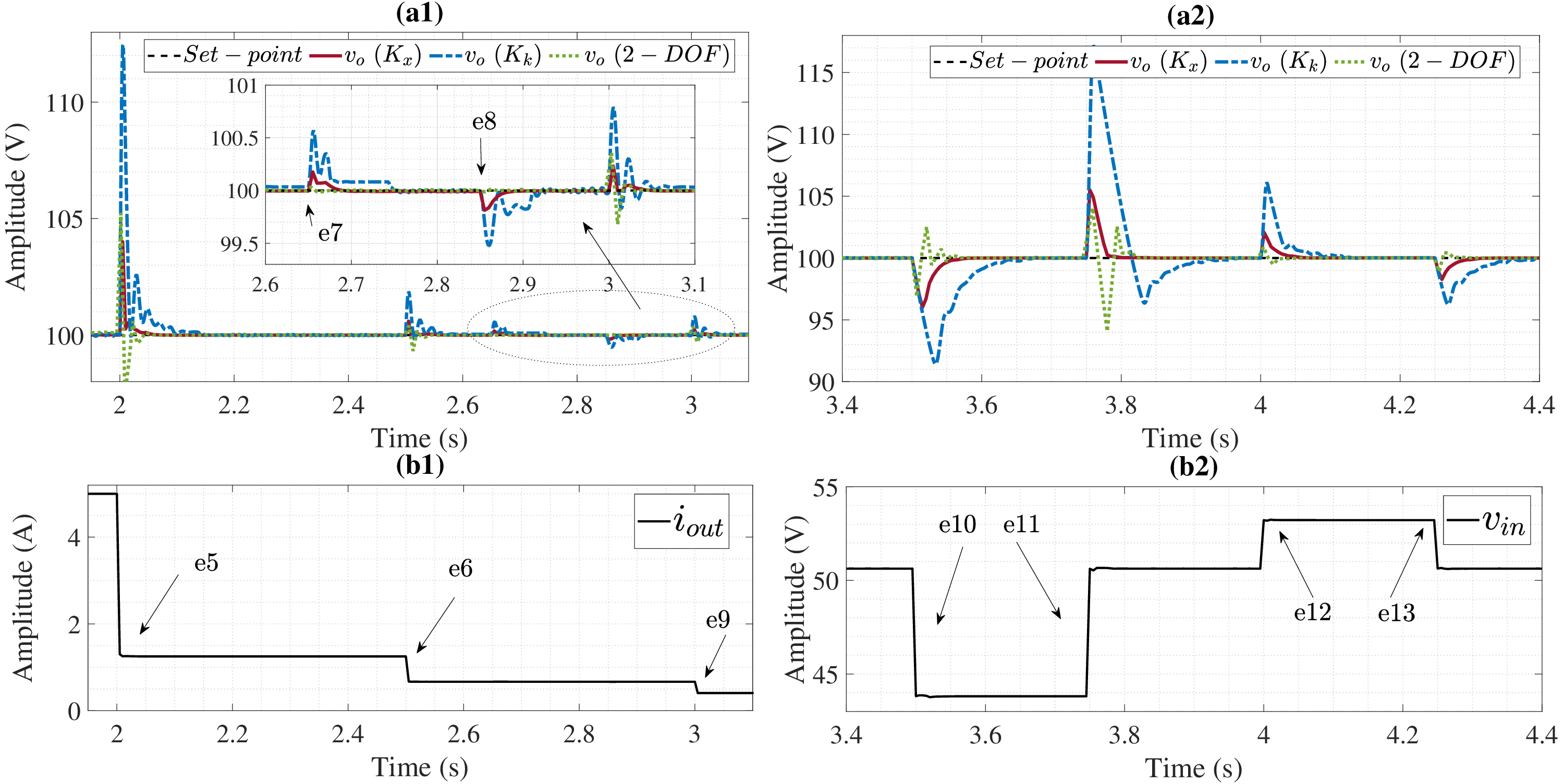}
\caption{The performance results of the control systems under the problematic events: Eliminating the effect of parameter uncertainties on the output voltage (a1, b1) Point 3: Step changing of the load resistance and the inductances ($v_{in}= 50$ V), (a2, b2) Point 4: The input voltage changes ($i_{out}= 0.4$ A)}
\label{fig6}
\end{figure}

\begin{table}
\caption{Performance indexes of the compensators (event 1-13=e1-13)}
    \centering
    \begin{tabular}{ccccccc}
 \hline
  \multirow{2}{*}{Point}& \multicolumn{3}{c}{Max over. ($\%$)} & \multicolumn{3}{c}{Sett. time (ms)} \\
 & $K_{x}$ &$K_{k}$& $2DOF$& $K_{x}$ & $K_{k}$ & $2DOF$ \\
 \hline
  \hline
Starting (e1)     &  0  & 0 & 1.5& 110 & 285& 35   \\
Point1 (e2)    &  0  & 0& 0.1 & 75 & 180& 70\\
Point 2 (e3)   &  2.3  & 5.8& 1.5 & 90 & 162& 90  \\
Point 2 (e4)   &  2.29  & 5.77& 1.95 & 74 & 154& 74 \\
Point 3 (e5)    &  4  & 12.5& 5.05 & 60 & 150& 60  \\
Point 3 (e6)   &  0.6  & 1.8& 1.7 & 15 & 75& 15 \\
Point 3 (e7)    &  0.2  & 0.6& 0.1 & 25 & 100& 10 \\
Point 3 (e8)   &  0.19  & 0.52& 0.1 & 23 & 125& 10  \\
Point 3 (e9)   &  0.2  & 0.5& 0.6 & 24 & 125& 30 \\
Point 4 (e10)   &  3.99  & 8.58& 3 & 80 & 180& 80  \\
Point 4 (e11)   &  5.5  & 17.3& 4.5 & 55 & 230& 55  \\
Point 4 (e12)   &  2.1  & 6.1& 0.85 & 65 & 140& 50   \\
Point 4 (e13)   &  1.73  & 3.75& 0.9 & 60 & 130 & 50\\
\hline
    \end{tabular}
    \label{tab3}
\end{table}
 Table 2 shows that the proposed control systems provide better robustness than the other compensator in the presence of variations of all parameters. The proposed control systems mitigate the oscillations under disturbance effects and provide solid tracking and uncertainty rejection performance under variations of the circuit parameters in terms of overshoot and settling time performance metrics. By adding disturbance loops to the synthesis process, output voltage oscillations are prevented in the presence of parametric uncertainties. The results demonstrate that, as expected, the proposed convex method outperforms the K-factor-based control strategy in the case of set-point tracking, disturbance, and uncertainty rejection performances.

\section{Experimental results}
The proposed control systems are implemented in a TMS320F28335 DSP-based experimental setup to validate the practical performance of the method. The general experimental setup and the gate driver circuit are demonstrated in Figs. 9 and 10. The PWM blocks of the DSP are used to realize 180 degrees of phase shifting. The MOSFETs of the converter are IRF-1310N with 36 $m\si{\ohm}$ conduction resistance; the passive switches are SB5100 Schottky diodes with negligible resistances. The TLP-350 optocouplers are used for gate isolation. Fig. 11 demonstrates the start-up of the converter and the set-point tracking performance with the control input signal (duty cycle of the PWM signals).
\begin{figure}
    \centering
    \includegraphics[width=7cm]{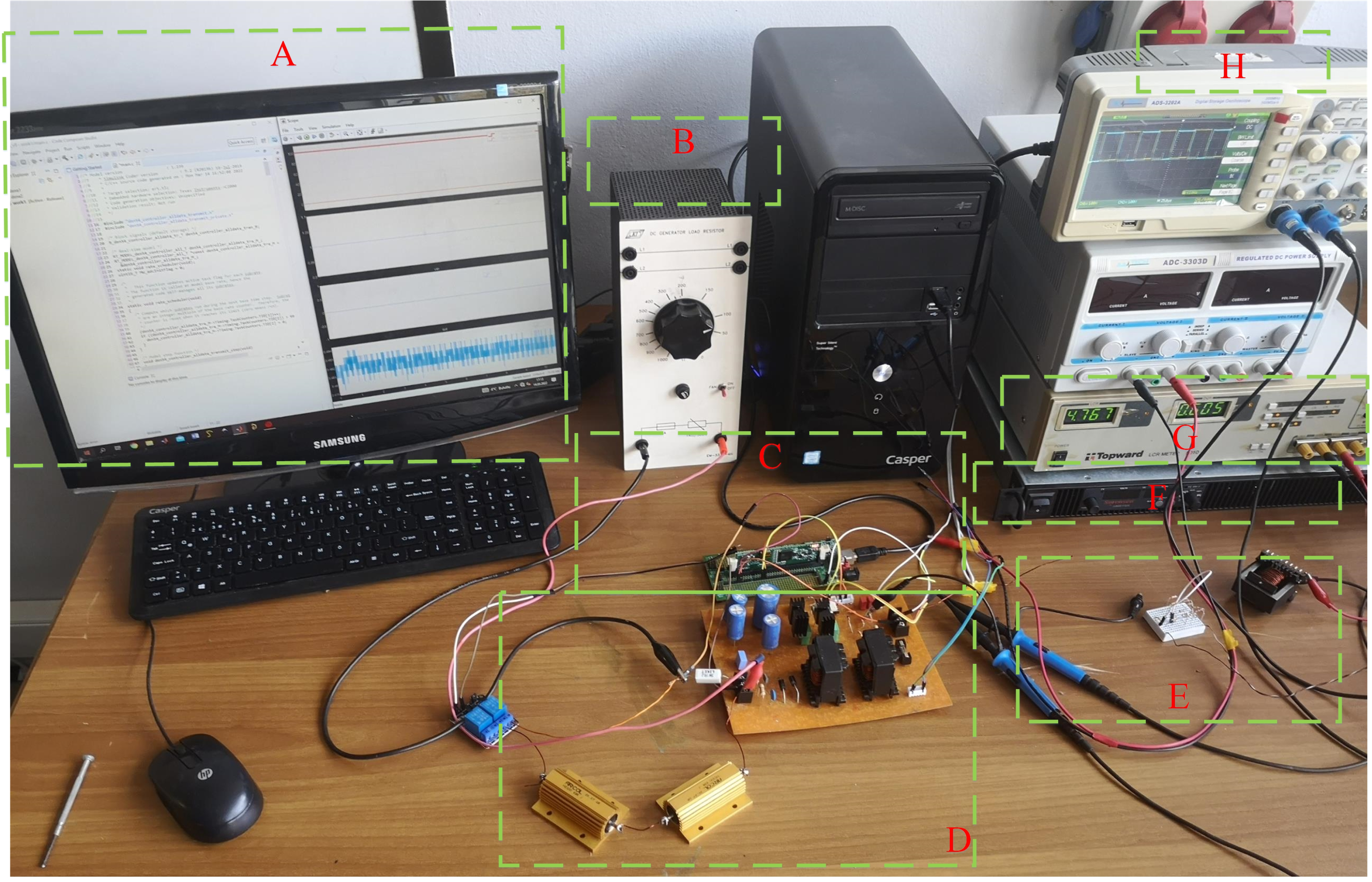}
    \caption{Experimental test bench: (A) host computer, (B) load resistor, (C) TMS320F28335 DSP, (D) 2-phase interleaved boost converter, (E) switching connection of the Sorensen programmable power supply, (F) Sorensen XG 300-5 power supply, (G) LCR meter, (H) ADS-3202A oscilloscope to monitor the PWM signals}
    \label{fig6}
\end{figure}
\begin{figure}
    \centering
    \includegraphics[width=8cm]{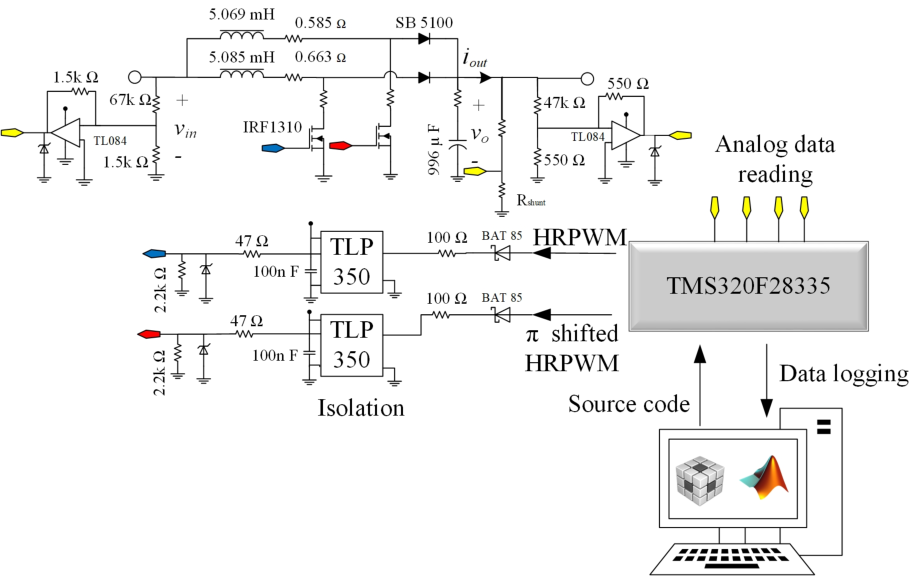}
    \caption{The electronic application and drive circuits of the control system}
    \label{fig6}
\end{figure}
Firstly, the starting process of the 1-DOF control system is shown in Fig. 11 (a), where the input voltage and load resistance are set to 52 V and 87 $\Omega$. The set-point value is changed between 22-26 seconds to analyze the set-point tracking performance in detail. The controlled system has no overshoots or undershoots at the starting point. It has a nearly 180 ms settling time similar to the simulation study presented in Fig. 7 (a1). Set-point tracking performance is good at rising edge changes of the set-point signal although an overshoot occurs on the falling edges of the set-point signal as in the simulation study presented in Fig. 6 (a). There is no steady-state error in the presence of the sensor noise and the disturbance effects. Fig. 12 demonstrates output voltage responses under the input voltage and output resistance changes. To ensure a constant output voltage, the current and input voltage amplitudes are changed over a certain period in Fig. 12. The control system mitigates the voltage oscillations on the output voltage of the converter which is exposed to input voltage changes. The converter maintains its stability and performance over a wide input voltage range. The circuit is subjected to a wide range of load changes as shown in Fig. 12 (b). The tracking performance of the 2-DOF controlled system is demonstrated in Fig. 13. The system has nearly no overshoots at set-point changes. The load current rejection performance of the 2-DOF system is presented to a wide range of load changes in Fig. 12 (b). The disturbance rejection performance of the 2-DOF system is presented under the load current and input voltage changes in Fig. 14. The output regulation is achieved after each load and input voltage change. In step load changes, the overshoots in the output voltage are quite low. The control systems provide solid performance above the rated load and in no-load operations. The experimental performances demonstrate that the proposed 1-DOF and 2-DOF control systems provide satisfactory dynamic performance with regard to output voltage tracking and robustness to load and input voltage changes. 
\begin{figure}
\centering
\subfloat[Starting of the converter: The output voltage (a1) and control input/duty cycle (a2)]
{\includegraphics[width=7cm]{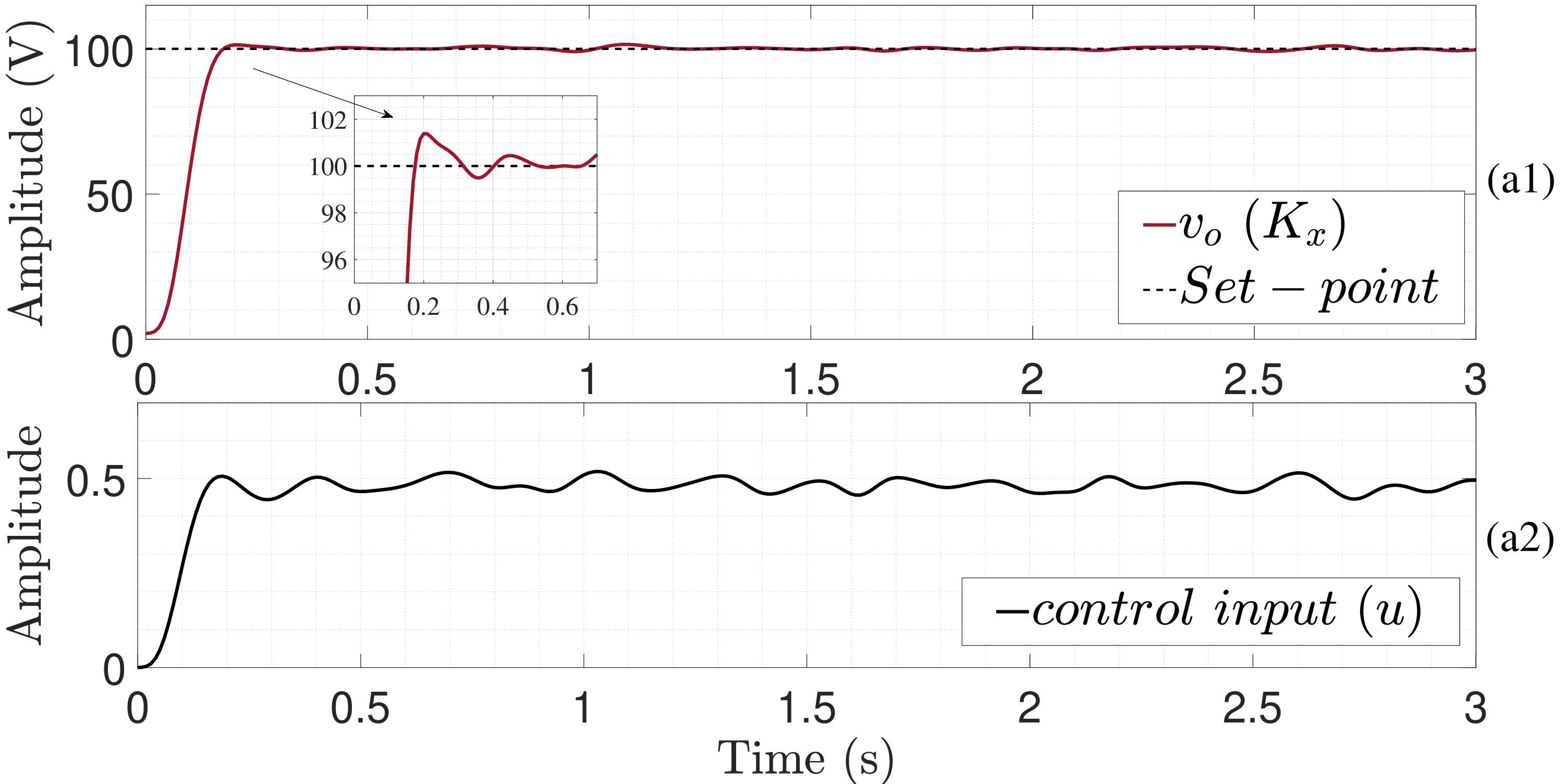}
\label{ber_vs_PS}} \\
\subfloat[Set-point tracking performance: The output voltage (b1) and control input (b2) ]{\includegraphics[width=7cm]{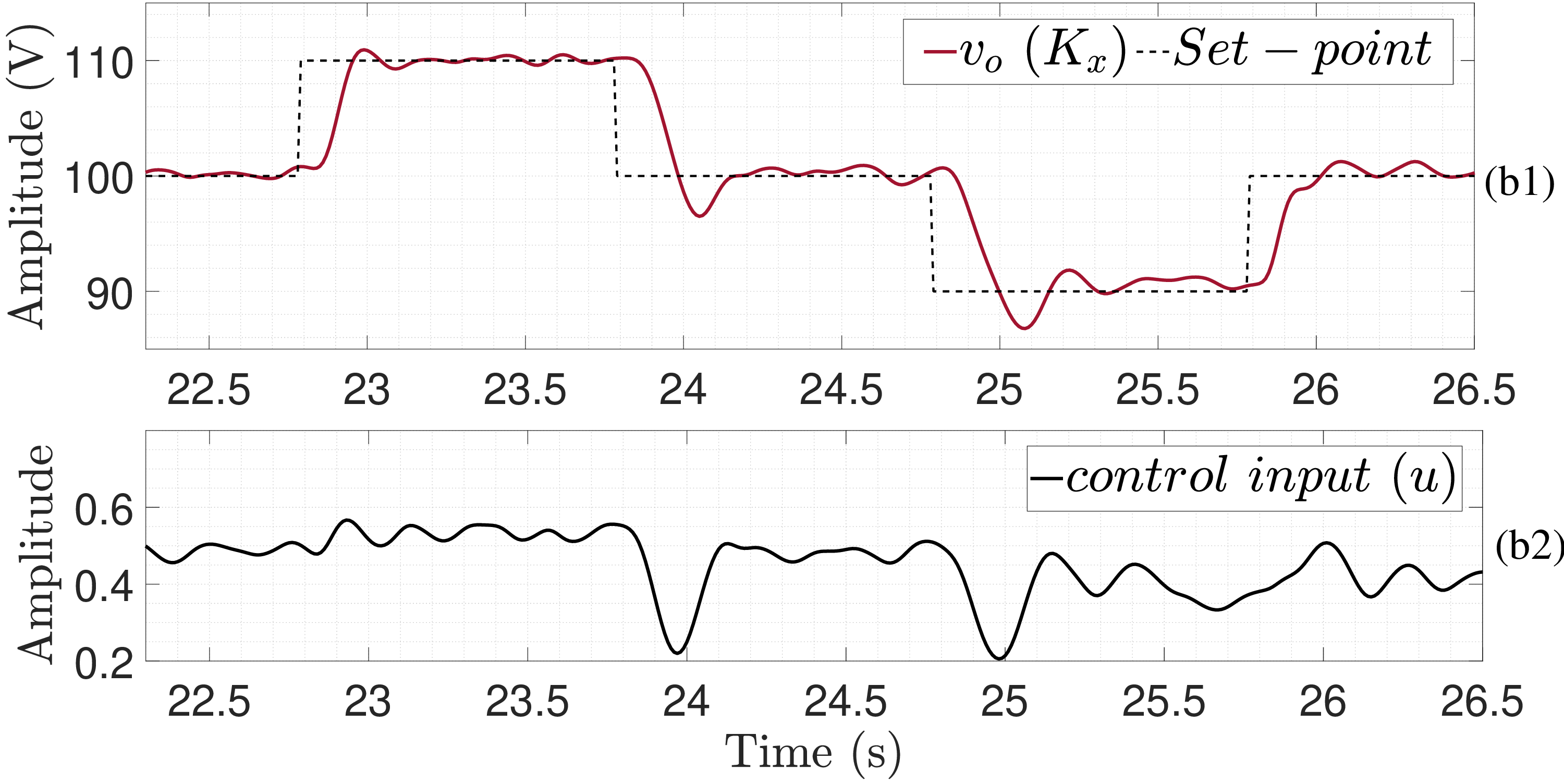}
\label{ber_vs_TS}} 
\caption{Real-time application results of the proposed compensator in the case of set-point changing: the output voltage and control input}
\label{fig8}
\end{figure}

\begin{figure}
\centering
\subfloat[Input voltage changes: The output voltage (a1) and input voltage (a2) ]{\includegraphics[width=7cm]{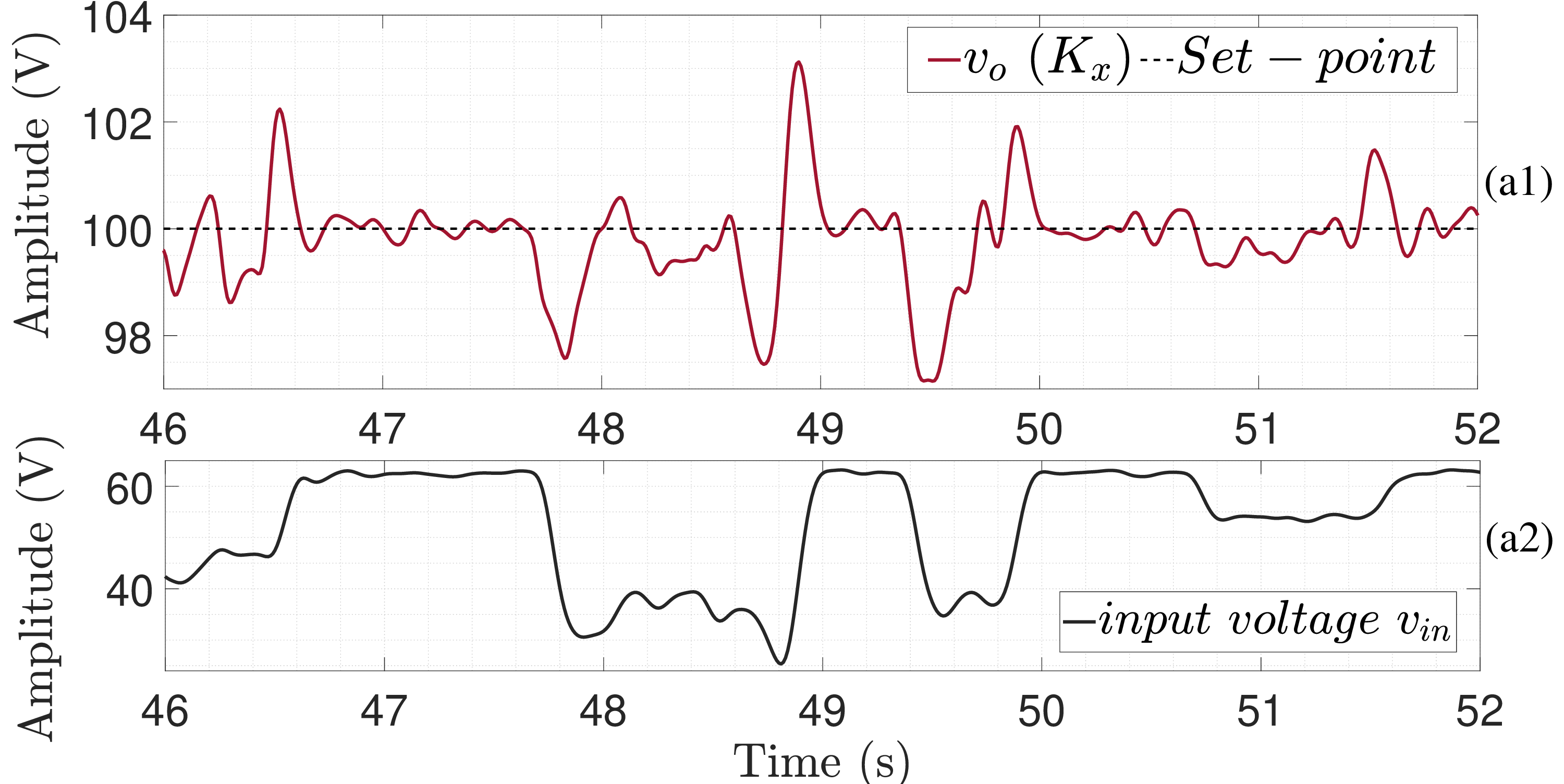}
\label{ber_vs_PS}} \\
\subfloat[Load changes: The output voltage (b1) and load current (b2)]{\includegraphics[width=7cm]{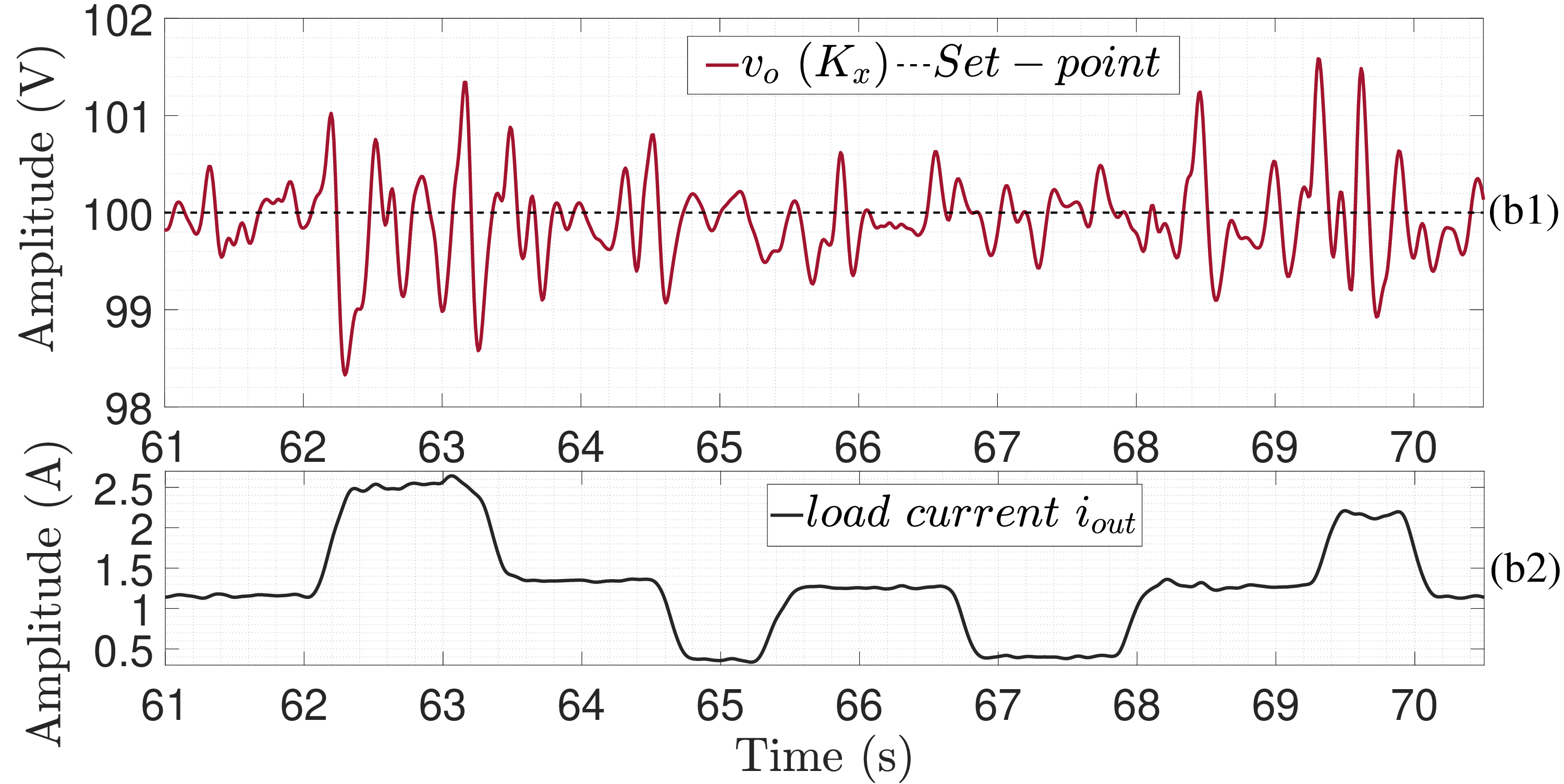}
\label{ber_vs_PS}} 
\caption{Real-time application results of the proposed compensator under crucial input voltage and load disturbances: the output voltage and control input}
\label{fig6}
\end{figure}

\begin{figure}
\centering
\subfloat{\includegraphics[width=7cm]{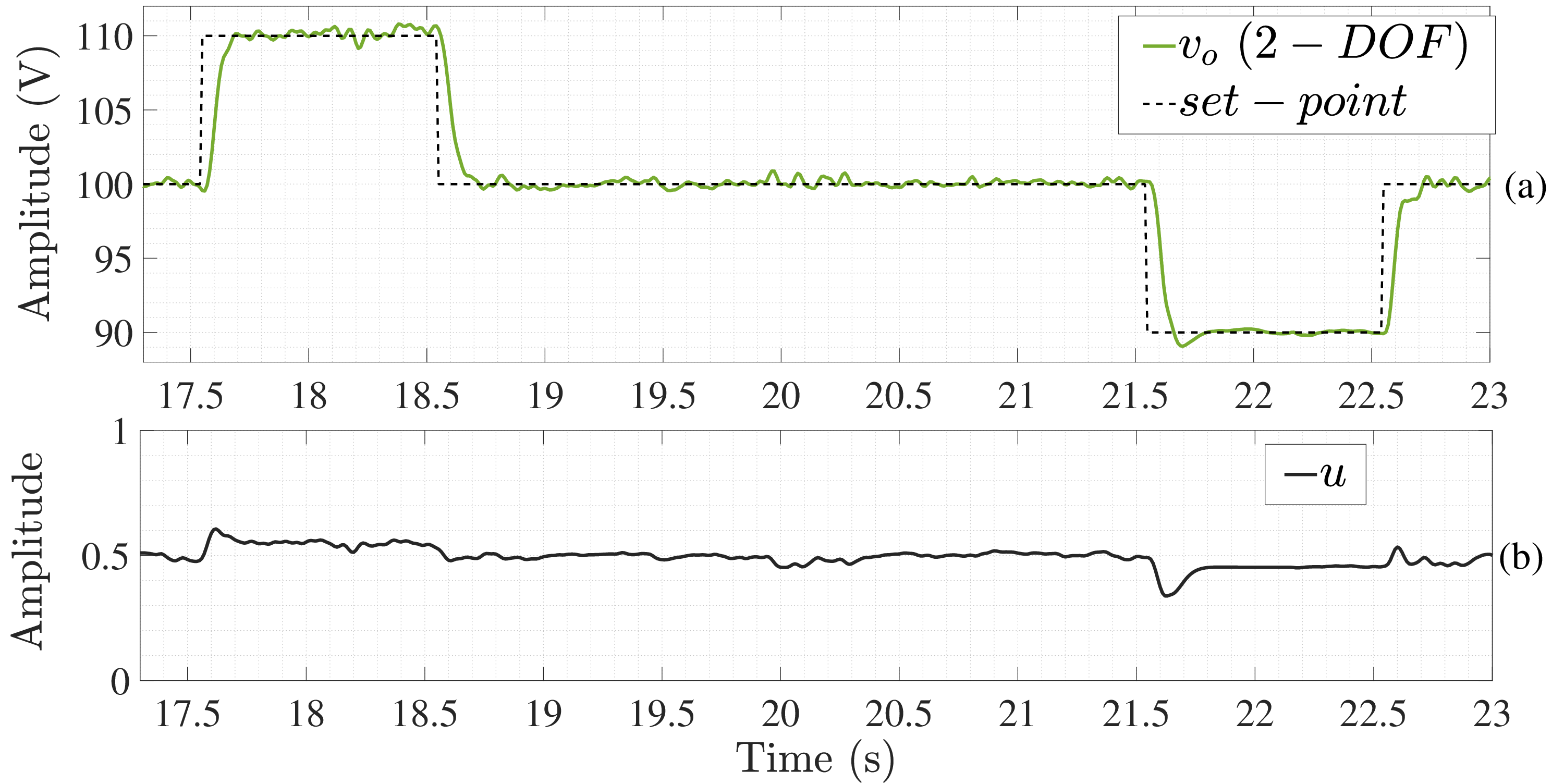}
\label{ber_vs_TS}} 
\caption{Real-time application results of the proposed 2-DOF control system in the case of set-point changes: the output voltage and control input}
\label{fig8}
\end{figure}

\begin{figure}
\centering
\subfloat[Step input voltage changes: The output voltage and input voltage]{\includegraphics[width=7cm]{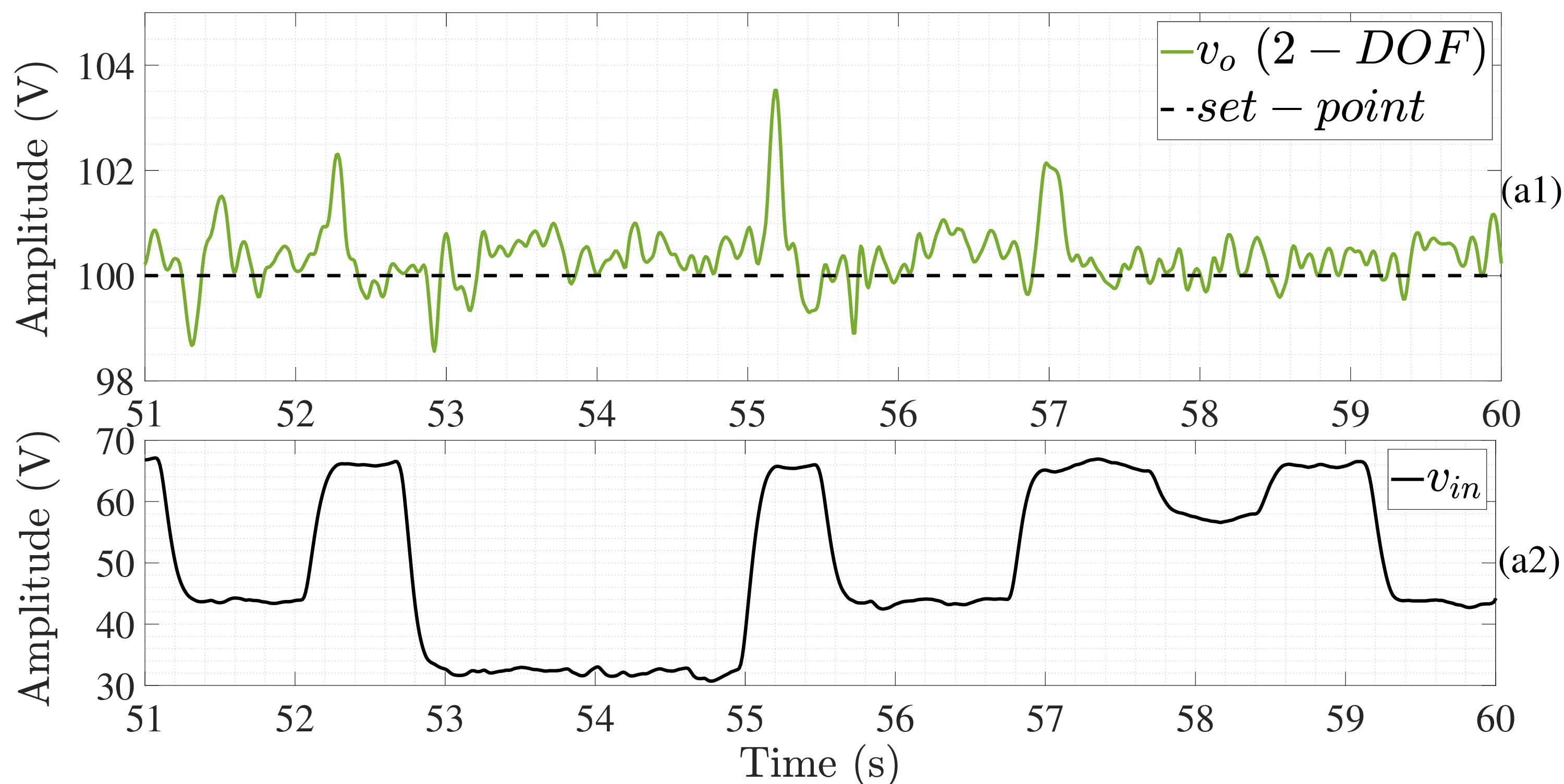}
\label{ber_vs_PS}} \\
\subfloat[Load changes: The output voltage and load current]{\includegraphics[width=7cm]{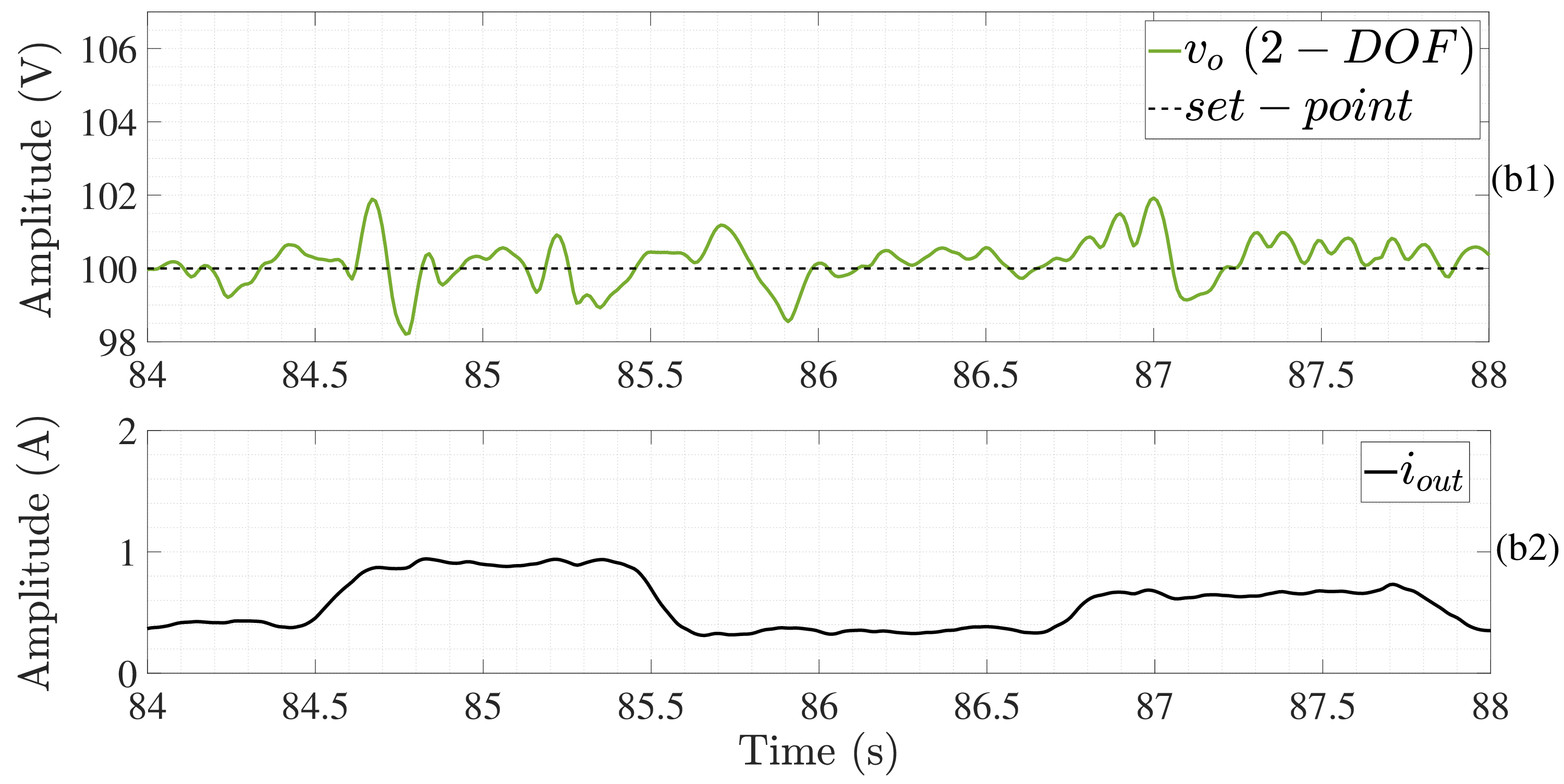}
\label{ber_vs_PS}} 
\caption{Real-time application results of the proposed 2-DOF control systems under input voltage and load disturbances}
\label{fig6}
\end{figure}
The slight differences between simulation and experiment results in Figs. 7, 8 and 11, 12 are due to minor differences in the disturbances. It is important to note that such a disconformity is due to the difference in sensitivity of the programmable power supplies in simulation and experiments, and the inevitable errors implied in practice by the measuring sensors. Also, heating of the passive components or sensor noise can cause such minor mismatches.

\section*{Discussion and Conclusion}
This paper proposes a simultaneous $H_\infty$ Type-III compensator and DOB synthesis approach for the multi-phase DC-DC power converters in a convex optimization framework. The proposed $H_\infty$ control schematic involves significant sensitivity functions to systematically take into account all relevant system constraints. It simplifies the design of error, control input, and noisy output weighting functions. We have proposed an iterative LMI algorithm to solve the design of the compensator and filter simultaneously. The proposed optimization problem is solved with several iterations in which the compensator is chosen as a unit integrator at the initial of the problem. Although the variation of the input voltage is significant (± 33$\%$) in the experimental setup, the output voltage of the 1-DOF controlled system has a small deviation of 200 ms and returns to the nominal set-point value without oscillations. The load current is changed between +25$\%$ and -75$\%$  of the nominal value. The output voltage presents small overshoots less than 1.5$\%$ of the nominal value during the load variations. The proposed 2-DOF control system is capable of rejecting input voltage and load current disturbances. Consequently, the effectiveness and performance of the proposed approaches against the parameter uncertainty and low-high load operation are validated in simulation and real-time applications. It is shown that the simulation and experimental results are consistent. The proposed  2-DOF control system is robust to disturbances and sensor noise.
\backmatter
\bibliographystyle{unsrt}
\bibliography{1_rkeskin}

\begin{thebibliography}{10}

\bibitem{azarastemal2021cascade}
Sajad Azarastemal and Mohammad Hejri.
\newblock Cascade control system design and stability analysis for a dc--dc
  boost converter with proportional integral and sliding mode controllers and
  using singular perturbation theory.
\newblock {\em Iranian Journal of Science and Technology, Transactions of
  Electrical Engineering}, 45(4):1445--1462, 2021.

\bibitem{shruti2021analytical}
P~Shruti, YG~Praveen, CP~Vipin, and B~Chitti Babu.
\newblock Analytical tuning of 2-dof smith predictor control scheme for dc-dc
  boost converter: A process control perspective.
\newblock {\em International Journal of Circuit Theory and Applications},
  49(3):641--655, 2021.

\bibitem{sarrafan2020novel}
Neda Sarrafan, Jafar Zarei, Roozbeh Razavi-Far, Mehrdad Saif, and
  Mohammad-Hassan Khooban.
\newblock A novel on-board dc/dc converter controller feeding uncertain
  constant power loads.
\newblock {\em IEEE Journal of Emerging and Selected Topics in Power
  Electronics}, 9(2):1233--1240, 2020.

\bibitem{hayes2016design}
Brendan Hayes, Marissa Condon, and Damian Giaouris.
\newblock Design of pid controllers using filippov's method for stable
  operation of dc--dc converters.
\newblock {\em International Journal of Circuit Theory and Applications},
  44(7):1437--1454, 2016.

\bibitem{chen2020interleaved}
Shin-Ju Chen, Sung-Pei Yang, Chao-Ming Huang, and Yu-Hua Chen.
\newblock Interleaved high step-up dc--dc converter with voltage-lift and
  voltage-stack techniques for photovoltaic systems.
\newblock {\em Energies}, 13(10):2537, 2020.

\bibitem{ghosh2016design}
Arnab Ghosh, Subrata Banerjee, Mrinal~Kanti Sarkar, and Priyanka Dutta.
\newblock Design and implementation of type-ii and type-iii controller for
  dc--dc switched-mode boost converter by using k-factor approach and
  optimisation techniques.
\newblock {\em IET Power Electronics}, 9(5):938--950, 2016.

\bibitem{rana2017development}
Niraj Rana, Arnab Ghosh, and Subrata Banerjee.
\newblock Development of an improved tristate buck--boost converter with
  optimized type-3 controller.
\newblock {\em IEEE Journal of Emerging and Selected Topics in Power
  Electronics}, 6(1):400--415, 2017.

\bibitem{anzehaee2018augmenting}
Mohammad~Mousavi Anzehaee, Behzad Behnam, and Payman Hajihosseini.
\newblock Augmenting armarkov-pfc predictive controller with pid-type iii to
  improve boost converter operation.
\newblock {\em Control Engineering Practice}, 79:65--77, 2018.

\bibitem{keskin2019design}
Ridvan Keskin and Ibrahim Aliskan.
\newblock Design of non-inverting buck-boost converter for electronic ballast
  compatible with led drivers.
\newblock {\em Karaelmas Science and Engineering}, 8(2):473--481, 2018.

\bibitem{cr2011type}
CR~CR.
\newblock Type iii compensator design for power converters.
\newblock {\em Power Electronics}, 2011.

\bibitem{banerjee2016improved}
Subrata Banerjee, Arnab Ghosh, and Niraj Rana.
\newblock An improved interleaved boost converter with pso-based optimal
  type-iii controller.
\newblock {\em IEEE Journal of Emerging and Selected Topics in Power
  Electronics}, 5(1):323--337, 2016.

\bibitem{tran2020dual}
Dai-Duong Tran, Sajib Chakraborty, Thomas Geury, Joeri Van~Mierlo, Mohamed
  El~Baghdadi, and Omar Hegazy.
\newblock Dual-loop control scheme with optimized type-iii controller based on
  genetic algorithm for 6-phase interleaved converter in electric vehicle
  drivetrains.
\newblock In {\em 2020 22nd European Conference on Power Electronics and
  Applications (EPE'20 ECCE Europe)}, pages P--1. IEEE, 2020.

\bibitem{rana2021performance}
Niraj Rana, Santosh Sonar, and Subrata Banerjee.
\newblock Performance investigation of closed loop dual phase interleaved buck
  boost converter with dragonfly optimized type iii controller.
\newblock {\em IEEE Transactions on Circuits and Systems II: Express Briefs},
  2021.

\bibitem{hindi2004tutorial}
Haitham Hindi.
\newblock A tutorial on convex optimization.
\newblock In {\em Proceedings of the 2004 American Control Conference},
  volume~4, pages 3252--3265. IEEE, 2004.

\bibitem{keskin2021robust}
R{\i}dvan Keskin, Ibrahim Aliskan, and Ersin Da{\c{s}}.
\newblock Robust structured controller synthesis for interleaved boost
  converters using an $h_\infty$ control method.
\newblock {\em Transactions of the Institute of Measurement and Control}, page
  01423312211019560, 2021.

\bibitem{sedhom2020robust}
Bishoy~E Sedhom, Magdi~M El-Saadawi, Ahmed~Y Hatata, Mostafa~A Elhosseini, and
  Elhossaini~E Abd-Raboh.
\newblock Robust control technique in an autonomous microgrid: A multi-stage
  $h\infty $ controller based on harmony search algorithm.
\newblock {\em Iranian Journal of Science and Technology, Transactions of
  Electrical Engineering}, 44(1):377--402, 2020.

\bibitem{bagheri2022robust}
Tahereh Bagheri~Rouch and Ahmad Fakharian.
\newblock Robust control of islanded dc microgrid for voltage regulation based
  on polytopic model and load sharing.
\newblock {\em Iranian Journal of Science and Technology, Transactions of
  Electrical Engineering}, 46(1):171--186, 2022.

\bibitem{ahmad2017robust}
Nur~Syazreen Ahmad.
\newblock Robust stability analysis and improved design of phase-locked loops
  with non-monotonic nonlinearities: Lmi-based approach.
\newblock {\em International Journal of Circuit Theory and Applications},
  45(12):2057--2072, 2017.

\bibitem{rodriguez2020modeling}
Armando~A Rodriguez, Karan Puttannaiah, Kaustav Mondal, Shiba Biswal, and Brent
  Wallace.
\newblock Modeling and control of a flapping wing hawkmoth micro air vehicle
  using generalized mixed sensitivity hierarchical design approach.
\newblock In {\em AIAA Scitech 2020 Forum}, page 2073, 2020.

\bibitem{erol2019fixed}
Bilal Erol and Ak{\i}n Deliba{\c{s}}{\i}.
\newblock Fixed-order $h_\infty$ controller design for mimo systems via
  polynomial approach.
\newblock {\em Transactions of the Institute of Measurement and Control},
  41(7):1985--1992, 2019.

\bibitem{ankelhed2012partially}
Daniel Ankelhed, Anders Helmersson, and Anders Hansson.
\newblock A partially augmented lagrangian method for low order h-infinity
  controller synthesis using rational constraints.
\newblock {\em IEEE transactions on automatic control}, 57(11):2901--2905,
  2012.

\bibitem{kammer2017decentralized}
Christoph Kammer and Alireza Karimi.
\newblock Decentralized and distributed transient control for microgrids.
\newblock {\em IEEE Transactions on Control Systems Technology},
  27(1):311--322, 2017.

\bibitem{dinh2011combining}
Quoc~Tran Dinh, Suat Gumussoy, Wim Michiels, and Moritz Diehl.
\newblock Combining convex--concave decompositions and linearization approaches
  for solving bmis, with application to static output feedback.
\newblock {\em IEEE Transactions on Automatic Control}, 57(6):1377--1390, 2011.

\bibitem{saeki2010low}
Masami Saeki, Masashi Ogawa, and Nobutaka Wada.
\newblock Low-order $h_\infty$ controller design on the frequency domain by
  partial optimization.
\newblock {\em International Journal of Robust and Nonlinear Control:
  IFAC-Affiliated Journal}, 20(3):323--333, 2010.

\bibitem{shinoda2017multivariable}
Shogo Shinoda, Kazuhiro Yubai, Daisuke Yashiro, and Junji Hirai.
\newblock Multivariable controller design achieving diagonal dominance using
  frequency response data.
\newblock {\em Electronics and Communications in Japan}, 100(10):12--23, 2017.

\bibitem{boyd2016mimo}
Stephen Boyd, Martin Hast, and Karl~Johan {\AA}str{\"o}m.
\newblock Mimo pid tuning via iterated lmi restriction.
\newblock {\em International Journal of Robust and Nonlinear Control},
  26(8):1718--1731, 2016.

\bibitem{segovia2013noise}
Vanessa~Romero Segovia, Tore H{\"a}gglund, and Karl~Johan {\AA}str{\"o}m.
\newblock Noise filtering in pi and pid control.
\newblock In {\em 2013 American Control Conference}, pages 1763--1770. IEEE,
  2013.

\bibitem{mercader2016robust}
Pedro Mercader, Karl~Johan {\AA}str{\"o}m, Alfonso Ba{\~n}os, and Tore
  H{\"a}gglund.
\newblock Robust pid design based on qft and convex--concave optimization.
\newblock {\em IEEE transactions on control systems technology},
  25(2):441--452, 2016.

\bibitem{kammer2018convex}
Christoph Kammer, Salvatore D’Arco, Atsede~Gualu Endegnanew, and Alireza
  Karimi.
\newblock Convex optimization-based control design for parallel grid-connected
  inverters.
\newblock {\em IEEE Transactions on Power Electronics}, 34(7):6048--6061, 2018.

\bibitem{karimi2017data}
Alireza Karimi and Christoph Kammer.
\newblock A data-driven approach to robust control of multivariable systems by
  convex optimization.
\newblock {\em Automatica}, 85:227--233, 2017.

\bibitem{ohnishi1996motion}
Kouhei Ohnishi, Masaaki Shibata, and Toshiyuki Murakami.
\newblock Motion control for advanced mechatronics.
\newblock {\em IEEE/ASME transactions on mechatronics}, 1(1):56--67, 1996.

\bibitem{sariyildiz2019disturbance}
Emre Sariyildiz, Roberto Oboe, and Kouhei Ohnishi.
\newblock Disturbance observer-based robust control and its applications: 35th
  anniversary overview.
\newblock {\em IEEE Transactions on Industrial Electronics}, 67(3):2042--2053,
  2019.

\bibitem{wang2004design}
Chun-Chih Wang and Masayoshi Tomizuka.
\newblock Design of robustly stable disturbance observers based on closed loop
  consideration using $h_\infty$ optimization and its applications to motion
  control systems.
\newblock In {\em Proceedings of the 2004 American Control Conference},
  volume~4, pages 3764--3769. IEEE, 2004.

\bibitem{wang2020robust}
Lu~Wang and Jianhua Cheng.
\newblock Robust disturbance rejection methodology for unstable non-minimum
  phase systems via disturbance observer.
\newblock {\em ISA transactions}, 100:1--12, 2020.

\bibitem{tena2022performance}
David Tena, Ignacio Pe{\~n}arrocha-Al{\'o}s, and R~Sanchis.
\newblock Performance, robustness and noise amplification trade-offs in
  disturbance observer control design.
\newblock {\em European Journal of Control}, 65:100630, 2022.

\bibitem{huang2022high}
Wei-Wei Huang, Peng Guo, Chuxiong Hu, and Li-Min Zhu.
\newblock High-performance control of fast tool servos with robust disturbance
  observer and modified $h_\infty$ control.
\newblock {\em Mechatronics}, 84:102781, 2022.

\bibitem{wang2022frequency}
Xiaoke Wang, Wataru Ohnishi, and Takafumi Koseki.
\newblock Frequency response data based disturbance observer design: With
  application to a nonminimum phase motion stage.
\newblock {\em IEEE/ASME Transactions on Mechatronics}, 27(6):5318--5326, 2022.

\bibitem{gonccalves2022disturbance}
Eduardo~Nunes Gon{\c{c}}alves and Denise~Fonseca Pereira.
\newblock Disturbance-observer-based control for multivariable systems based on
  equivalent transfer function.
\newblock {\em Journal of Control, Automation and Electrical Systems},
  33(6):1700--1710, 2022.

\bibitem{keskin2021multi}
R{\i}dvan Keskin, Ibrahim Aliskan, and Ersin Da{\c{s}}.
\newblock Multi-variable modeling and system identification of an interleaved
  boost converter.
\newblock In {\em 2021 13th International Conference on Electrical and
  Electronics Engineering (ELECO)}, pages 550--554. IEEE, 2021.

\bibitem{chan2015generalized}
Ka-Fai Chan, Chi-Seng Lam, Wen-Liang Zeng, Wen-Ming Zheng, Sai-Weng Sin, and
  Man-Chung Wong.
\newblock Generalized type iii controller design interface for dc-dc
  converters.
\newblock In {\em TENCON 2015-2015 IEEE Region 10 Conference}, pages 1--6.
  IEEE, 2015.

\bibitem{keskin2023robust}
Rıdvan Keskin, Ibrahim Aliskan, and Ersin Da{\c{s}}.
\newblock Robust fixed-order $h_\infty$ controller synthesis for multi-phase
  converters with input voltage feed-forward.
\newblock {\em Journal of the Franklin Institute}, 2023.

\bibitem{campi2018introduction}
Marco~C Campi and Simone Garatti.
\newblock {\em Introduction to the scenario approach}.
\newblock SIAM, 2018.

\bibitem{dacs2021robust}
Ersin Daş and Selahattin~{\c{C}}a{\u{g}}lar Ba{\c{s}}lam{\i}{\c{s}}l{\i}.
\newblock Robust data-driven fixed-order controller synthesis: Model matching
  approach.
\newblock {\em IET Control Theory $\&$ Applications}, 2021.

\bibitem{dacs2021data}
Ersin Da{\c{s}} and Selahattin {\c{C}}a{\u{g}}lar~Ba{\c{s}}lam{\i}{\c{s}}l{\i}.
\newblock Data-driven fixed-order $h_\infty$ controller synthesis in frequency
  domain: Closed-loop system approach.
\newblock {\em Transactions of the Institute of Measurement and Control},
  43(5):1059--1067, 2021.

\bibitem{lofberg2008modeling}
Johan L{\"o}fberg.
\newblock Modeling and solving uncertain optimization problems in yalmip.
\newblock {\em IFAC Proceedings Volumes}, 41(2):1337--1341, 2008.

\end{thebibliography}

\end{document}